# A virtual instrument to standardise the calibration of atomic force microscope cantilevers

**John E. Sader[1*], Riccardo Borgani[2], Christopher T. Gibson[3], David B. Haviland[2], Michael J. Higgins[4], Jason I. Kilpatrick[5], Jianing Lu[6], Paul Mulvaney[6], Cameron J. Shearer[3], Ashley D. Slattery[3], Per-Anders Thorén[2], Jim Tran[1], Heyou Zhang[6], Hongrui Zhang[4], and Tian Zheng[4]**

[1]*School of Mathematics and Statistics, The University of Melbourne, Victoria, 3010, Australia*

[2]*Nanostructure Physics, Royal Institute of Technology (KTH), Roslagstullsbacken 21, SE-10691 Stockholm, Sweden*

[3]*Flinders Centre for NanoScale Science and Technology, School of Chemical and Physical Sciences, Flinders University, Bedford Park, South Australia 5042, Australia*

[4]*ARC Centre of Excellence for Electromaterials Science, Australian Institute for Innovative Materials, University of Wollongong, Wollongong, NSW 2522, Australia*

[5]*Conway Institute of Biomolecular and Biomedical Research, University College Dublin, Belfield, Dublin 4, Ireland*

[6]*School of Chemistry & Bio21 Institute, The University of Melbourne, Victoria, 3010, Australia*

[*] Corresponding author. E-mail: jsader@unimelb.edu.au

# ABSTRACT

Atomic force microscope (AFM) users often calibrate the spring constants of cantilevers using functionality built into individual instruments. This is performed without reference to a global standard, which hinders robust comparison of force measurements reported by different laboratories. In this article, we describe a virtual instrument (an internet-based initiative) whereby users from all laboratories can instantly and quantitatively compare their calibration measurements to those of others – standardising AFM force measurements – and simultaneously enabling non-invasive calibration of AFM cantilevers of any geometry. This global calibration initiative requires no additional instrumentation or data processing on the part of the user. It utilises a single website where users upload currently available data. A proof-of-principle demonstration of this initiative is presented using measured data from five independent laboratories across three countries, which also allows for an assessment of current calibration.

# I. INTRODUCTION

Standardisation of measurements is vital in metrology, allowing for quantitative comparison of data originating from different laboratories and users. While internationally accepted standards exist for many quantities, such as distance, time and mass, measurements that utilise force pose a particular challenge. This issue is especially relevant to the atomic force microscope (AFM) that uses a force-sensing probe, a microfabricated cantilever, to image surfaces and measure forces with atomic resolution.

The existing capability for users to calibrate the spring constant of individual AFM cantilevers is a highly utilised feature of modern commercial AFMs. This is often performed *in situ* with the cantilever loaded into the AFM using a number of methods; see Ref. [1-4] for reviews. The requirement for calibration stems from well-known variability in the material and dimensional properties (particularly thickness) of commercially available cantilevers. These variations can lead to departures from a manufacturer's nominally specified spring constant by an order-of-magnitude.

Based on early reports of these calibration methods and more recent research that include an inter-laboratory study [1-7], error estimates for their operation have been formulated. Such (averaged) estimates are typically used to quantify the accuracy of individual force measurements, but do not account for the variation of measurement uncertainty between different users and laboratories. This hinders direct comparison of AFM force measurements between laboratories.

One calibration method commonly built into commercial AFMs monitors the Brownian fluctuations of a cantilever and makes use of the equipartition theorem to determine its spring constant – the "AFM Thermal method" [8]. While applicable to any cantilever, spring constant measurements from this method are known to depend on several factors including the laser spot size and position, $z$-displacement piezo calibration, static-to-dynamic optical lever sensitivity and non-linearity in the deflection curve on hard contact [1-6, 9, 10]. Thus, variations in calibration accuracy can easily emerge between laboratories and even different users within a single laboratory. A round-robin study has reported such variations previously [5], a finding that we independently assess here – large variations on individual cantilevers are also found, see below.

The current absence of a readily accessible international force standard thus provides impetus to develop alternative approaches to standardise force measurements made using the AFM.

The aim of this article is to describe an internet-based virtual instrument, which enables all users in the AFM community to instantly compare and standardise their spring constant calibration measurements. This calibration initiative utilises a live database, accessed via a single website, where users can upload their AFM Thermal method measurements in air of the spring constant, resonant frequency and quality factor of a given cantilever type. By combining and appropriately averaging this dataset through use of a recent theoretical development of the "Sader method", described in Refs. [11-13], uncertainty in the measured spring constant is systematically reduced. This averaging process intrinsically accommodates true variations in the spring constant between cantilevers that arise from variability in their material properties and thickness. Internet-based collation enables the large number of independent calibration measurements made in laboratories across the world to be correlated and statistically significant measurements assessed.

The averaging process of this calibration initiative naturally generates the hydrodynamic function of that cantilever type, embodied in (what we term here) its $A$-coefficient, see Section II [11]. This immediately enables non-invasive calibration of AFM cantilevers (of any geometry) using the Sader method [11] – the method requires only the cantilever's resonant frequency and quality factor in air. The global averaging of data from many users eliminates the need to independently determine the $A$-coefficient of each cantilever type as described in Refs. [11-13].

The validity of this internet-based approach is assessed using a three-stage methodology. First, a single set of 10 cantilevers is calibrated using a laser Doppler vibrometer (LDV) to provide accurate benchmark data [6, 11]. Second, the same set of cantilevers is calibrated using the AFM Thermal method, by 5 separate groups involving 11 users (5 different universities and 3 countries). This is performed with no group knowing the identities of the other groups until data collection is complete. This constitutes a blind implementation of the AFM Thermal method between groups – as commonly occurs in calibration measurements from different laboratories. Strikingly, strong variations in the measured spring constants of the 10 individual cantilevers are observed (up to a factor greater than four), see Section IV-A-2. Even so, averaging this dataset appropriately [13] reproduces the $A$-coefficient obtained using LDV. Third, these groups calibrate unknown cantilevers of the same type, which they procure independently, using the AFM Thermal method – this is the intended operation of the calibration initiative. Averaging this dataset recovers precisely the same $A$-coefficient as the other two approaches.

This three-stage validation shows that the standard AFM Thermal method can be used with confidence to determine the $A$-coefficient of a particular cantilever type via the calibration website

– using cantilevers procured by anyone in the AFM community. This gives users the ability to immediately compare their calibration measurements to those of others in the AFM community, while systematically refining their own calibration using others' data. It thus provides a reference point for worldwide calibration, enabling force measurements from different laboratories to be compared directly.

An important observation of the present study is that the AFM Thermal method can generate highly precise measurements of the resonant frequency and quality factor of an individual cantilever – in contrast to the spring constant, discussed above. This is because complicating factors that affect measurement of the spring constant using the AFM Thermal method (such as laser spot position, spot size, and $z$-piezo calibration) do not influence the measured resonant frequency and quality factor. The technical requirements for precision measurements of the resonant frequency and quality factor of any cantilever are minimal and easily accommodated by the user, see Section IV-A-2. This measurement precision, and user-independence, motivates the present initiative whereby global standardisation of spring constant calibration is founded on measurement of the resonant frequency and quality factor only.

This Global Calibration Initiative (GCI), its internet platform (a single website [14]) and modes of operation are explained here together with results of the above-mentioned proof-of-principle implementation. The aim is to demonstrate the GCI and its ability to standardise AFM spring constant calibration. Its development serves to (i) build confidence in the accuracy and precision of force measurements made in any one laboratory, and (ii) allow the wide series of discrete datasets around the world to be combined and generate much more useful and significant measurement sets. Vitally, the GCI utilises existing equipment and measurements, and thus requires no extra effort on the part of the user – readily available data from the AFM Thermal method are uploaded to the GCI website [14], stored confidentially and used to systematically refine calibration; see Section III.

## II. THEORETICAL FRAMEWORK

The GCI makes use of both the AFM Thermal method and the Sader method. The AFM Thermal method (which is built into many commercial instruments) provides the raw calibration data that individual users collect, whereas the theoretical framework of the Sader method is used to refine the accuracy of this dataset. By using both methods together, the necessary $A$-coefficient is determined for accurate application of the Sader method alone, for any cantilever type, as discussed above.

The Sader method relates the measured resonant frequency and quality factor of a cantilever, measured in air (typically), to its spring constant. In its original form, the method was formulated for cantilevers of rectangular geometry [15]. This was subsequently generalised to accommodate cantilevers of arbitrary geometry [16] – the hydrodynamic function defines a universal dimensionless function for a particular cantilever type, i.e., plan view geometry; see Section II of Ref. [11]. With knowledge of this hydrodynamic function for a given geometry, the spring constant of a cantilever is evaluated from measurement of its resonant frequency and quality factor in air. A number of approaches have been formulated to determine this hydrodynamic function [11, 16].

Here, we make use of a recent theoretical development of the Sader method that greatly simplifies determination of the hydrodynamic function; described in Section III-D of Ref. [11]. This relates the resonant frequency, $f_{R,\mathrm{ref}}$, quality factor, $Q_{\mathrm{ref}}$, and spring constant, $k_{\mathrm{ref}}$, of a reference cantilever to the same parameters, $(f_R, Q, k)$, of an uncalibrated cantilever of identical plan view dimensions:

$$k = k_{\mathrm{ref}} \frac{Q}{Q_{\mathrm{ref}}} \left( \frac{f_R}{f_{R,\mathrm{ref}}} \right)^{2-\alpha}, \tag{1}$$

where $\alpha = 0.7$ [11, 12]. This result is weakly dependent on variations in plan view dimensions [11], the accuracy of which has been discussed and verified in several recent studies [12, 13, 17, 18]. This formula has most recently been generalised to accommodate multiple (different) reference cantilevers, through appropriate averaging, leading to the required formula [13]:

$$k = A\, Q {f_R}^{1.3}, \tag{2}$$

where the $A$-coefficient is universal for a particular cantilever geometry, and given by

$$A = \frac{1}{N} \sum_{i=1}^{N} A_i = \frac{1}{N} \sum_{i=1}^{N} \frac{k_{\mathrm{ref},i}}{Q_{\mathrm{ref},i}\, f_{R,\mathrm{ref},i}^{1.3}}, \tag{3}$$

where $N$ is the total number of measurements (each measurement shall henceforth be referred to as a “data-point”) and the subscript $i$ refers to an individual reference cantilever. The $A$-coefficient together with Eq. (2) implicitly defines the hydrodynamic function for a particular cantilever type.

Equation (3) provides the foundation for the standardisation algorithm of the GCI, enabling AFM Thermal method calibration measurements of multiple (and different) reference cantilevers to be compared and averaged together. This averaging formulation over multiple cantilevers systematically reduces uncertainty in the $A$-coefficient, resulting from the measured parameters of the reference cantilevers – typically dominated by that of the spring constant, as we demonstrate below. It also intrinsically accommodates any real variation in the spring constants, resonant frequencies and quality factors of individual cantilevers, providing a robust averaging methodology. The standardisation procedure of the GCI is described in the next section.

## III. DESCRIPTION OF THE GLOBAL CALIBRATION INITIATIVE

The GCI is accessed via a single website [14] where users upload data for the resonant frequency, quality factor and spring constant of a cantilever, measured using the AFM Thermal method (built into most commercial AFMs). As such, the GCI reports the spring constant directly at a cantilever's imaging tip position; eliminating the need for off-end correction. Measurements are always performed in air. Each cantilever type is allocated a separate database within the website because each type will generally have a different hydrodynamic function and hence $A$-coefficient; users suggest cantilever types which are then added to the GCI. Two modalities of operation are available depending on the number of data-points uploaded.

*Initialisation:* When a particular database for a given cantilever type is new, i.e., empty, or contains fewer than a fixed critical number of data-points (set to 20 in this proof-of-principle study – and indicated on the website), users can only upload their measured data. This collected data is averaged together using Eq. (3) giving an estimate for the $A$-coefficient and its standard error [13].

Once the minimum critical number of data-points is reached, users can then operate the website in two distinct ways:

Modality 1: Users input their measured resonant frequency, quality factor and spring constant obtained using the AFM Thermal method. This data-point is uploaded to the database, following which Eq. (3) is used to update the $A$-coefficient and its error estimate. A refined spring constant is calculated using the Sader method, Eq. (2), and reported. Users can then compare their measured spring constant (from the AFM

Thermal method) to this new value – derived from the measurements of all other users – providing an instant and live global calibration standard.

Modality 2: Users input their measured resonant frequency and quality factor only, again using the AFM Thermal method. The spring constant from the Sader method is then reported, using Eq. (2) and the current $A$-coefficient specified by Eq. (3). The standard error in the $A$-coefficient (calculated from the current database) is used to provide an up-to-date error estimate for the reported spring constant. No data is uploaded to the database and the $A$-coefficient is unchanged. This modality provides non-invasive calibration of the cantilever using the Sader method.

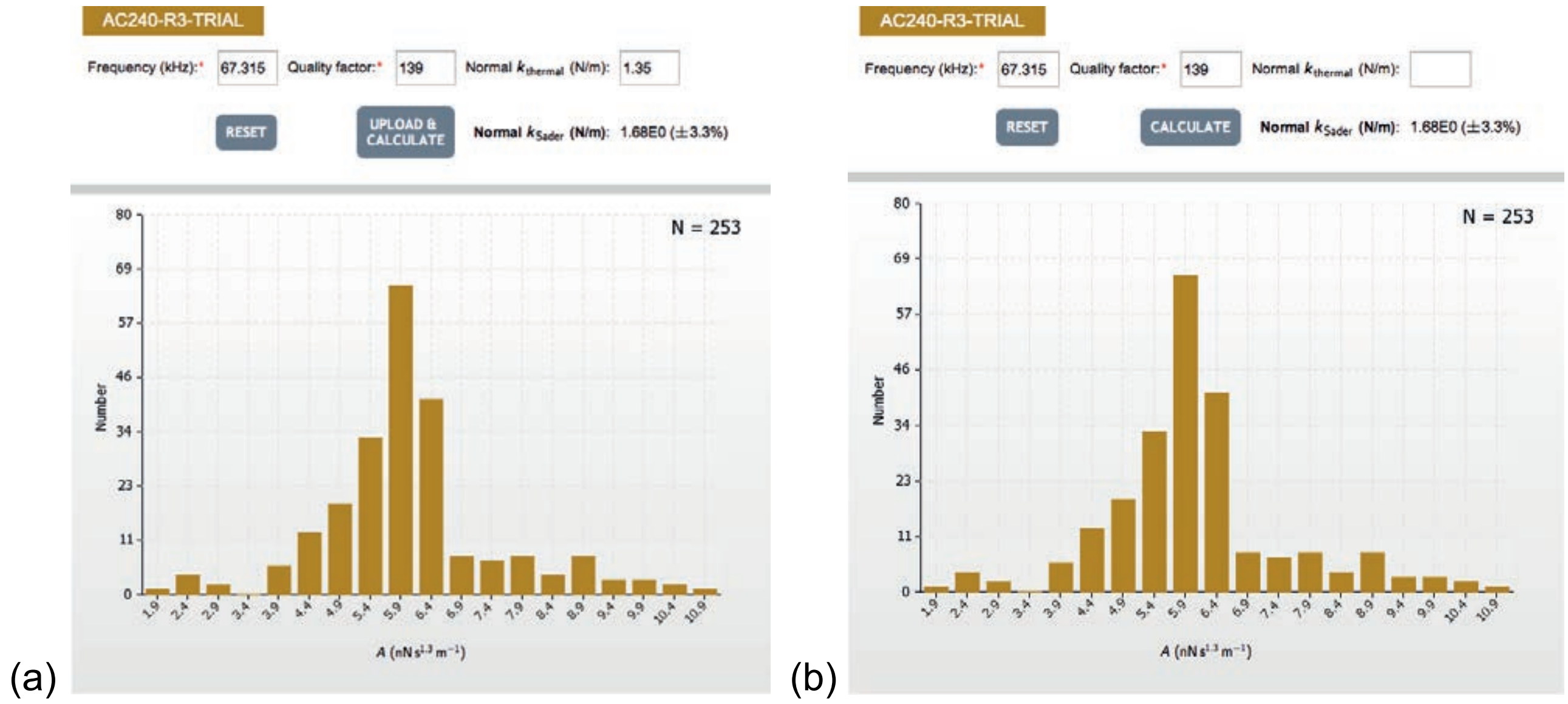

**FIG. 1:** Screen shots of the GCI showing operations in (a) modality 1, and (b) modality 2. These modalities are described in Section III and activated once the minimum number of data-points required is uploaded. Data is entered in the first row of the web interface. The current number of data-points is indicated in the histogram. Modality 1: Resonant frequency, quality factor and spring constant measured using the AFM Thermal method are input. "Upload & Calculate" button is activated which, upon selection, uploads the data-point and merges it with the current database. Equation (2) is used to determine the spring constant, which is returned with a 95% confidence interval based on the current database – enabling users to immediately compare and standardise their measurements to those of others. Modality 2: Resonant frequency and quality factor only are input and the "Calculate" button remains (the default). Pressing this button gives the spring constant and error estimate, as above. The data-point is not uploaded/merged with the current database.

The two modalities are selected automatically depending whether the spring constant, measured using the AFM Thermal method, is input or not. The first modality allows users to help each other, by providing a means to anonymously and robustly compare their calibration measurements to those of others while systematically refining the live estimate for the $A$-coefficient (which is the foundation for the GCI). The second modality makes use of this existing database to achieve non-invasive calibration. Screen shots of the GCI, as used in the proof-of-principle measurements of this study (reported in the next section), are given in Fig. 1.

In both modalities, a live cumulative histogram of individual $A$-coefficients determined from all users is displayed, also specifying the number of data-points and users in the database for that cantilever type – raw uploaded data is not revealed, see Fig. 1. Note that the number of users was not shown in the histogram during the proof-of-principle measurements (Fig. 1). Analysis of the database gives a measure of the uncertainty in the spring constant obtained using the AFM Thermal method, because uncertainty in the $A$-coefficient is dominated by that of the spring constant; see below.

*Data confidentiality:* The website and its data uploads are protected by SSL security via a freely available login/password unique to each user (which anyone in the AFM community can instantly gain). Each user can view their own uploaded data, which they can edit as needed, but users cannot view/edit the uploaded data of others. All uploaded data is stored for the sole purpose of analysis using the GCI and will not be disclosed to any third party – ensuring confidentiality.

## IV. PROOF-OF-PRINCIPLE IMPLEMENTATION

In this section, we report the results of a proof-of-principle implementation of the GCI involving five independent groups across three countries: Australia, Ireland and Sweden. To simulate operation of the GCI, which is available to all users in the AFM community, the 5 groups were not told of the each other's identities until they completed the data collection and uploaded their data to the website [14]. The groups could see the cumulative and current histogram for the $A$-coefficient during data collection, as shown in Fig. 1. No other data or information was available to them. This methodology facilitates a robust assessment of the GCI because individual users in the AFM community will not have access to each other's identities or raw data (i.e., resonant frequencies, quality factors and spring constants). The participants were also not given instructions on how to implement the AFM Thermal method, simulating any variability present in its current application. They were aware that their measured data would be compared at the end of the data collection period.

For this proof-of-principle assessment, Olympus AC240-R3 cantilevers were selected due to their widespread availability and use. The groups were asked to calibrate two sets of cantilevers:

A. A single set of ten AC240-R3 cantilevers that was passed sequentially to each of the 5 groups. These are labelled Cantilever 1, 2, 3, … , 10 for convenience.

B. Unspecified AC240-R3 cantilevers procured independently by 4 groups.

The first set enables direct comparison of measurements from different groups (and users) on identical cantilevers – eliminating any potential variability due to cantilever properties and thus providing a robust assessment of the GCI's theoretical framework. The latter mimics the intended operation of the GCI where users procure and calibrate their own cantilevers. Comparison and merging of data from these two sets allows the GCI's intended operation to be rigorously tested.

## A. Set of ten AC240-R3 cantilevers

We now present data for the first measurements, where all 5 participating groups calibrated a single set of ten AC240-R3 cantilevers. These cantilevers were procured from Bruker (OLTESPA-R3) and are rebranded Olympus AC240-R3 cantilevers. They are coated with a thin reflective Al coating. Since the Sader method is independent of such coatings [11], cantilevers with any coating can be used to determine the $A$-coefficient of a particular cantilever type. An optical image of one cantilever is given in Fig. 2.

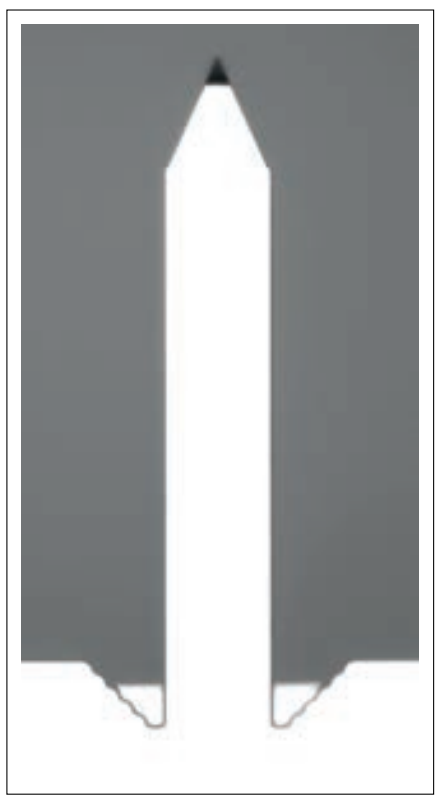

**FIG. 2:** Optical image of Cantilever 6 in its holding box. The imaging tip (black triangle) is pointing out of the page.

Calibration of this 10-cantilever set provides a strong measure of the AFM Thermal method's performance, as different users, groups and instruments are involved – they all calibrated exactly the same cantilevers. Such an assessment is critical because the AFM Thermal method supplies the raw data for the GCI. No cantilever was broken in this study, ensuring a full comparison of each cantilever across all groups and users.

### *1. Laser Doppler vibrometer measurements*

A laser Doppler vibrometer (LDV), MSA-500 Micro System Analyzer, Polytec (Waldbronn, Germany), was used to generate robust and accurate benchmark measurements for the resonant frequency, quality factor and spring constant of each cantilever in this 10-cantilever set. Brownian fluctuations of the cantilevers in air were measured at the laboratory temperature (17.5 °C).

Two approaches were used to calibrate each cantilever using the LDV, to ensure the robustness of measurements (discussed below):

Approach 1: Monitor the raw time series of the cantilever's Brownian motion (thermal fluctuations), and post-process this dataset to obtain its power spectral density (PSD), i.e., its thermal noise spectrum. This is as per Ref. [11].

Approach 2: Use the inbuilt PSD analysis of the MSA-500 to directly measure the thermal noise spectrum.

Since velocity is intrinsically calibrated in the LDV, use of the equipartition theorem enables the dynamic spring constant to be determined directly. This non-contact measurement differs strongly to the AFM Thermal method – the latter requires contact with a hard surface to calibrate the measured displacement, which can introduce significant uncertainty; see Section I. Finite element simulations were performed to convert the measured dynamic spring constant to a static spring constant, giving $k_{\mathrm{dyn}}/k_{\mathrm{stat}} = 1.051$ [10, 11, 19]. Static spring constants are reported here, as typically returned by the AFM Thermal method; see Refs. [10, 11] for a discussion.

The LDV measurement procedure enables uncertainties in the spring constants to be quantified. This is achieved by measuring the stiffness of the cantilever along a linear grid near its imaging tip and extrapolating to the tip position, as per Fig. 9 of Ref. [11]. As observed in that previous study [11], the resulting uncertainty varies between individual measurements and cantilevers; see below.

This uncertainty is due to scatter in the measured local stiffness along the linear grid, which is much larger than the uncertainty in fitting the measured PSDs (to obtain the local stiffness); see Ref. [11].

*a. Plan view dimensions* The plan view dimensions of all cantilevers were measured using the optical microscope in the MSA-500 LDV and a TEM grid as reference. They were found to be: *length* = 239.3±1.3(SD) µm, *width* = 39.3±0.2(SD) µm. The observed relative standard deviations in measured values of the widths and lengths are identical (0.5%). These measurements show that the plan view dimensions of the 10-cantilever set are highly uniform.

These 10 cantilevers are therefore ideally suited for use of Eqs. (2) and (3), which implicitly assume that the plan view dimensions of the reference and uncalibrated cantilevers [defined in Eq. (1)] are identical; this is discussed further below.

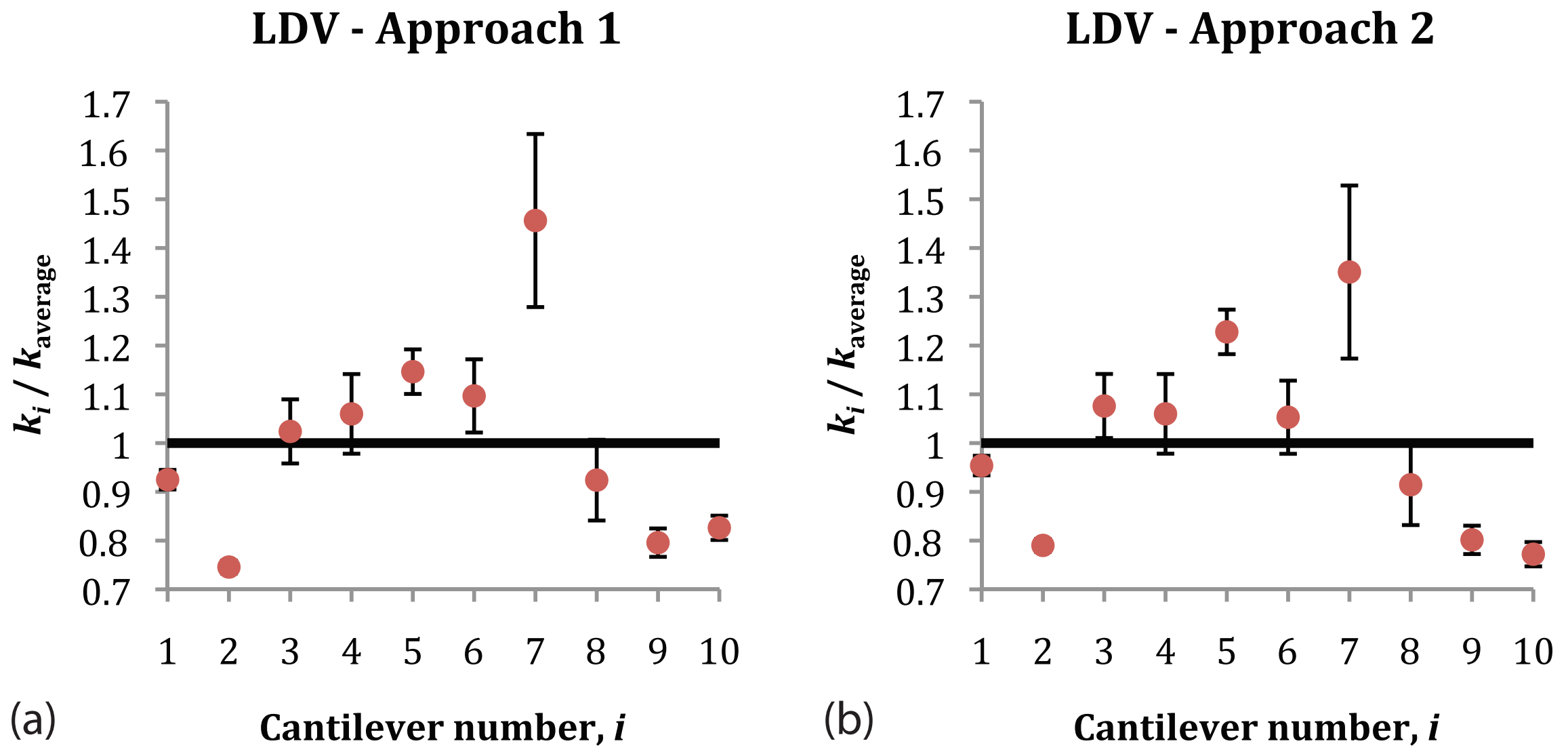

**FIG. 3:** LDV measured spring constant at the imaging tip position of the set of 10 cantilevers: (a) approach 1 (time series), and (b) approach 2 (direct PSD). Results are normalised by the average spring constant of all cantilevers. Subscript $i$ refers to the cantilever number, whereas "average" refers to the average over all $i$. Error bars are derived from fits to 5 independent measurements of stiffness versus position along each cantilever's axis. Some measurements display greater uncertainty than others (as observed in Ref. [11]). Error bars specify a 75% C.I.

*b. LDV spring constants* Figure 3 gives the LDV measured spring constants of each cantilever using the two above-stated approaches. The error bars specify a 75% confidence interval. Notably, of the 20 measurements only 5 coincide with the average spring constant – showing that the cantilevers exhibit significantly different stiffness despite their identical plan view dimensions (see above). This is also reflected in their measured resonant frequencies and quality factors, discussed below. The averaged spring constant of all 10 cantilevers obtained using the two approaches [i.e., in panels (a) and (b) of Fig. 3] are very similar, differing by only 1%, namely $k_{\text{ave}}^{(\text{a})} = 1.436$ N/m, $k_{\text{ave}}^{(\text{b})} = 1.422$ N/m. Both approaches are expected to produce identical results (barring statistical variation), and this observation supports that expectation.

*c. A-coefficient* Measurement of the PSD of thermal fluctuations of each cantilever immediately enables its resonant frequency, quality factor and spring constant to be determined (by fitting the PSD to the response of a simple harmonic oscillator, e.g., as per Refs. [5, 11, 22]). Critically, this measurement must be performed with the cantilever retracted well away from any surface to eliminate hydrodynamic squeeze film damping effects – which can strongly reduce the quality factor; this is discussed further in the next section. Because LDV measurements on all 10 cantilevers were made directly in the holding box supplied by the manufacturer, this condition was well satisfied.

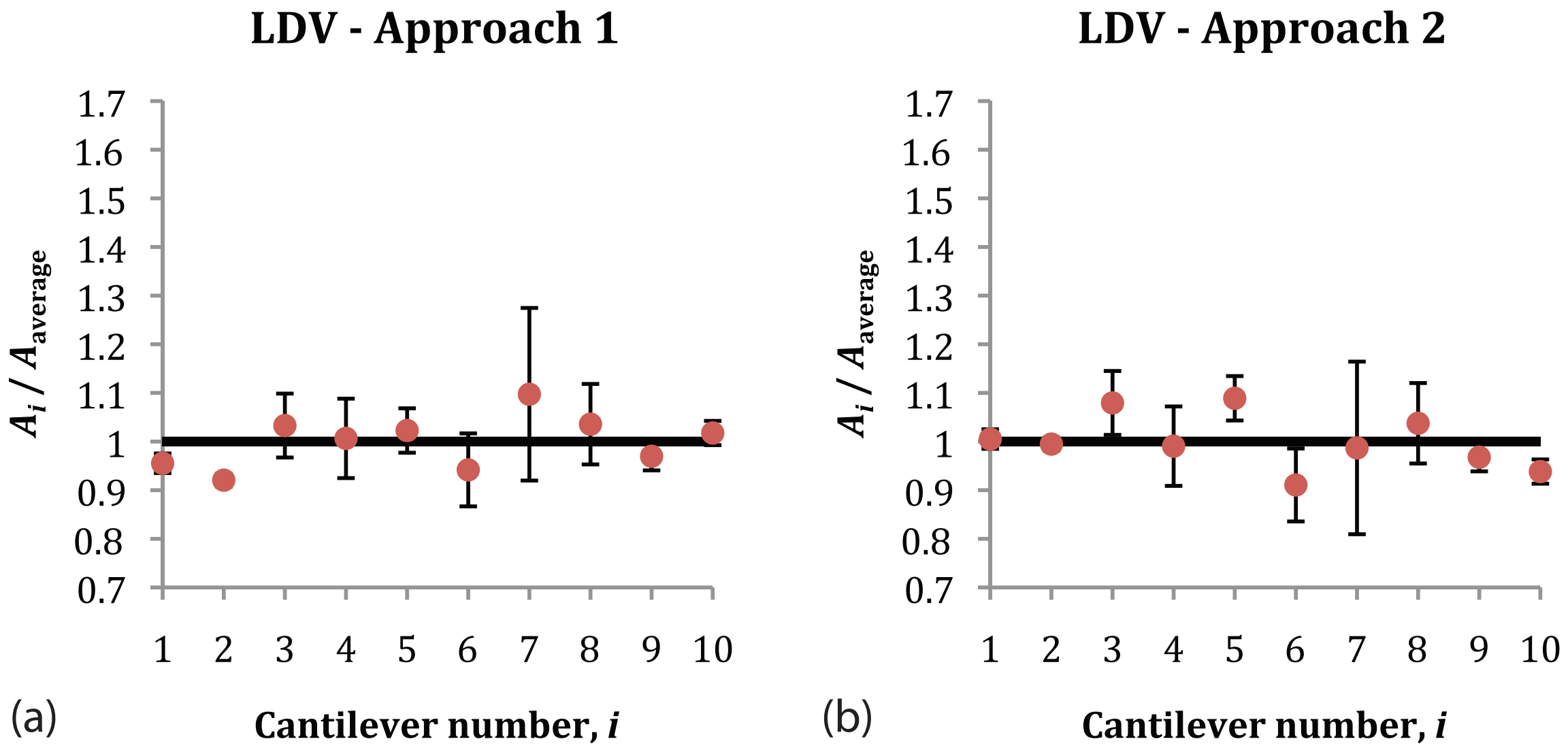

**FIG. 4:** LDV measured $A$-coefficients of set of 10 cantilevers: (a) approach 1 (time series), and (b) approach 2 (direct PSD). Results are normalised by the average $A$-coefficient of all cantilevers. Other details as per Fig. 3.

All 10 cantilevers are expected to yield identical results for the $A$-coefficient, because they are of the same type and have virtually identical plan view dimensions. Results for the $A$-coefficient are given in Fig. 4; horizontal and vertical scales identical to those in Fig. 3 are used to facilitate comparison. Again, error bars specify a 75% confidence interval. Strikingly, from a total of 20 independent measurements, error bars of 16 measurements encompass the average value (unity), i.e., 80% of measurements. This contrasts to the spring constant results in Fig. 3, where only 5 measurements encompass the mean (i.e., 25% of all measurements). Figures 3 and 4 thus demonstrate the robustness of the $A$-coefficient in normalising data from multiple cantilevers of the same type. The observed uncertainties are primarily due to those of the LDV measured spring constants [11].

Using Eq. (3), these measurements provide an $A$-coefficient for the AC240-R3 cantilevers of

$$A = 6.43 \pm 0.15 \; [\mathrm{nN\,s^{1.3}m^{-1}}] \,, \tag{4}$$

with the uncertainty in Eq. (4) now specifying a 95% confidence interval.

## 2. AFM Thermal Noise Measurements

The 5 participating Groups (11 different Users utilising a range of different AFMs) independently calibrated the 10-cantilever set. The AFMs used are Asylum Research Cypher-S, two separate Asylum Research MFP-3D, Bruker Dimension FastScan AFM, Bruker Icon with the Intermodulation Products Thermal method, Bruker Multimode 8 AFM with an E scanner, Bruker Multimode 8 AFM with a J scanner, JPK Biowizard II and JPK Nanowizard. As mentioned above, different users and groups performed their measurements independently and without knowledge of each other's results. Accuracy of these AFM measurements is assessed by comparison to the LDV benchmark data (previous section).

Details of the Groups and their Users are given in Table I. Note that some Groups consist of multiple users, whereas others have only one user that performs repeat measurements (which are allocated different User numbers).

Each cantilever was calibrated using the AFM Thermal method, yielding results for the resonant frequency, quality factor and spring constant (at the imaging tip position) of that cantilever. The AFM Thermal method is normally implemented in commercial AFMs using a dynamic-to-static conversion factor for the spring constant of 1.03, as per Refs. [10, 20], and thus it reports the static spring constant. Advanced calculations or finite element simulations are required to determine the true conversion factor for a particular cantilever type [11] – this was not implemented by the 5 participating groups to simulate general use by the AFM community. Groups 1, 2, 3 & 5 reported the static spring constant from the AFM Thermal method, whereas Group 4 reported the dynamic spring constant (a small difference, see above) which may also occur infrequently in world-wide use. In practice, the usual static spring constant from the AFM Thermal method (containing the factor of 1.03, above) should be uploaded to the GCI website for consistency.

**TABLE I:** Groups and Users participating in this study. "Unique" indicates an individual user. "Same, repeat" indicates that the one user in the Group performed repeat measurements that are assigned different User numbers. "Same, time series" indicates that a single user performed the LDV time-series measurements, whereas "Same, PSD" indicates that the same user performed the LDV PSD measurements – these datasets are allocated different User numbers.

| Group | User | Comments |
|---|---|---|
| LDV | 1 | Same, time series |
| | 2 | Same, PSD |
| 1 | 3 | Unique |
| | 4 | Unique |
| | 5 | Unique |
| 2 | 6 | Unique |
| | 7 | Unique |
| 3 | 8 | Same, repeat |
| | 9 | Same, repeat |
| 4 | 10 | Unique |
| | 11 | Unique |
| 5 | 12 | Unique |
| | 13 | Unique |

*a. Resonant frequency* Figure 5 gives a sample of measurements for the resonant frequency (of Cantilevers 7 and 10 – see Supplementary Information for other cantilevers, which behave similarly). LDV measurements are given by Users 1 and 2; all other Users report AFM data.

While there is some variation between the different Users (and Groups) – see Supplementary Information for expanded vertical axes – the observed variation in the measured resonant frequency for each cantilever is small (<0.05% SD). This variation is consistent with the expected uncertainty in determining the resonant frequency from the thermal noise spectrum [11, 21], and is two orders-of-magnitude smaller than the variation in resonant frequencies between the different Cantilevers 1 – 10 (6% SD). The LDV measurements were performed at a laboratory temperature of 17.5 °C, whereas the AFM measurement temperatures are expected to be higher (typically ~25 °C [22]) – temperatures in each AFM were not measured. This temperature difference will weakly affect the measurements and may explain the maximal values in the resonant frequency exhibited in the LDV data in Fig. 5; see Supplementary Information.

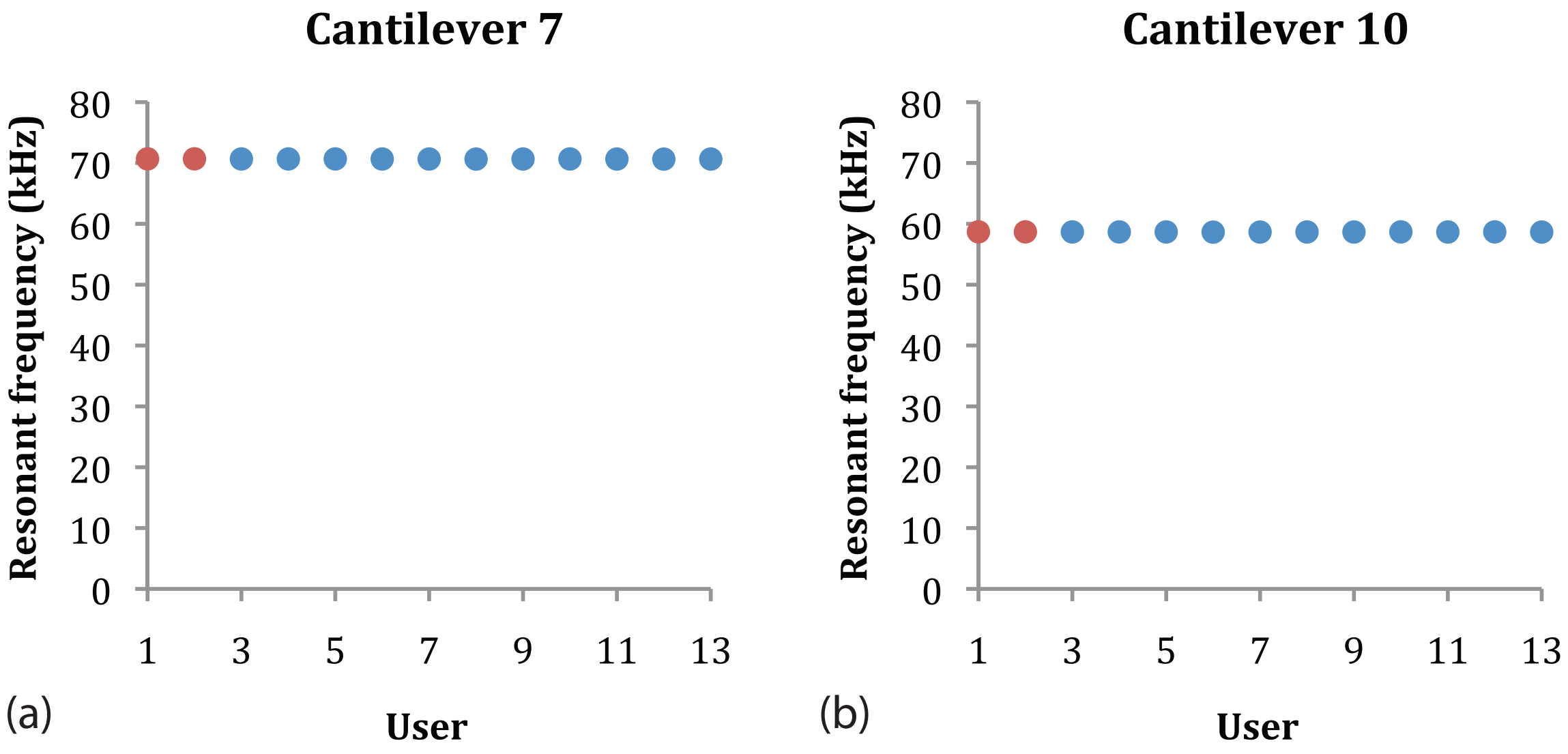

**FIG. 5:** Measured resonant frequencies for two of the cantilevers using the LDV (Users 1 & 2 – red) and AFM Thermal method (blue). Six Groups and 13 Users performed the measurements. Resonant frequency is expected to be constant as a function of User number – barring small variations due to statistical uncertainty and temperature variations – because all Users calibrated exactly the same cantilever. Identical vertical scales are used for comparison.

This analysis shows that all Groups precisely measure the resonant frequencies of all cantilevers using the AFM Thermal method.

*b. Quality factor* Figure 6 gives corresponding results for the quality factor (again, other cantilevers exhibit similar behaviour). In Fig. 6, all Users report comparable values with the exception of Users 11 & 13 whose data are systematically lower (by ~20%). These deviations are larger than the uncertainty expected from fitting the measured thermal noise spectrum (typically at the 1% level [21, 22]). Interestingly, while these Users frequently reported quality factors lower than those of other Users, this was not always the case – results for Cantilever 8 from Users 11 & 13 are more consistent with those of other Users (see Supplementary Information). Some other Users also occasionally report anomalously low values for the quality factors; see Supplementary Information.

It is known that bringing the cantilever in proximity to a solid surface will reduce the quality factor due to squeeze film damping. While Users 11 & 13 typically report low quality factors relative to other Users (in Fig. 6), User 11 is from Group 4, whereas User 13 is from Group 5.

These anomalous results are therefore most likely due to insufficient retraction of the cantilever from the surface, rather than an instrumentation issue. We also note that the resonant frequency is weakly affected by surface proximity in comparison to the quality factor [23, 24] (see Supplementary Information) – explaining why such an anomaly is not observed in the measured resonant frequencies in Fig. 5.

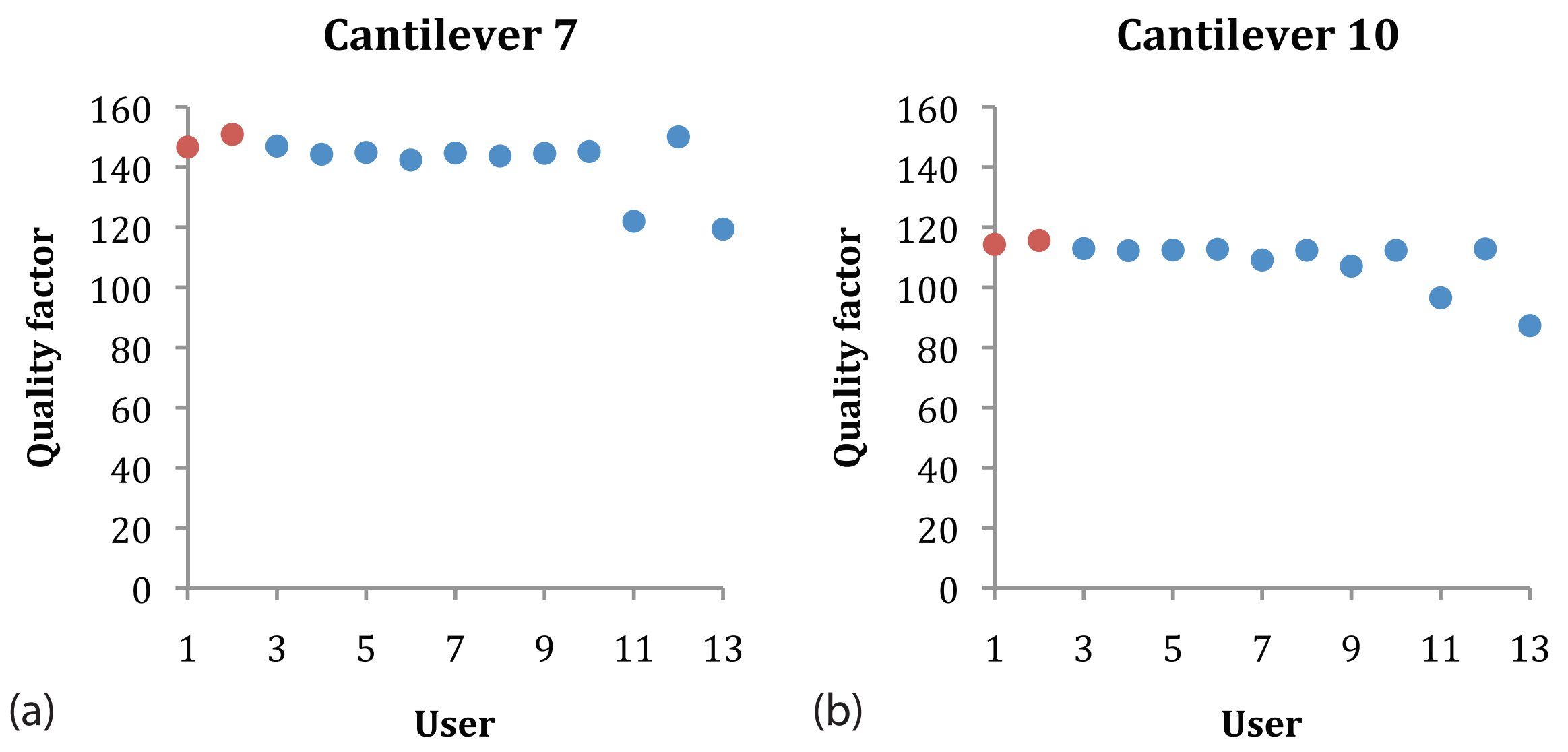

**FIG. 6:** Measured quality factors for two of the cantilevers using the LDV (Users 1 & 2 – red) and AFM Thermal method (blue). Six Groups and 13 Users performed the measurements. Quality factor is expected to be constant as a function of User number – barring small variations due to statistical uncertainty and temperature variations – because all Users calibrated exactly the same cantilever. Identical vertical scales are used for comparison.

The average relative SD in the measured quality factors of each cantilever, across all Users, is ~7%. Removing Users 11 and 13 from this calculation, reduces the value to 3% – consistent with the expected uncertainty in measuring quality factors from the thermal noise spectrum, see above. This highlights an important requirement:

> For an accurate measurement of the quality factor, the cantilever must be retracted sufficiently from the surface (by at least several cantilever widths) to curtail squeeze film damping by the surrounding air; see Supplementary Information for an example. With such operation, accurate data for the quality factor (and resonant frequency) can be ensured.

Finite frequency resolution in signal processing of the measured cantilever deflection can also lead to an (artificial) underestimate of the quality factor. This instrumentation issue can be easily corrected using Eq. (B1) of Ref. [11], if required.

*c. Spring constant* Figure 7 presents results for the spring constants of Cantilevers 7 and 10 measured using the AFM Thermal method; again, measurements of other cantilevers give comparable behaviour, see Supplementary Information. We emphasise that the AFM Thermal method intrinsically measures the spring constant at the imaging tip position. Unlike results for the resonant frequency and quality factor, the measured spring constant of a single cantilever is found to vary strongly between Users. While there are some trends that indicate that this may be instrumentation related (e.g., Users 6 & 7 are from the same Group, as are Users 10 & 11 – see Table I), strong variations also occur between Users of the same Group (e.g., Users 12 and 13). The reference LDV data are given by Users 1 and 2.

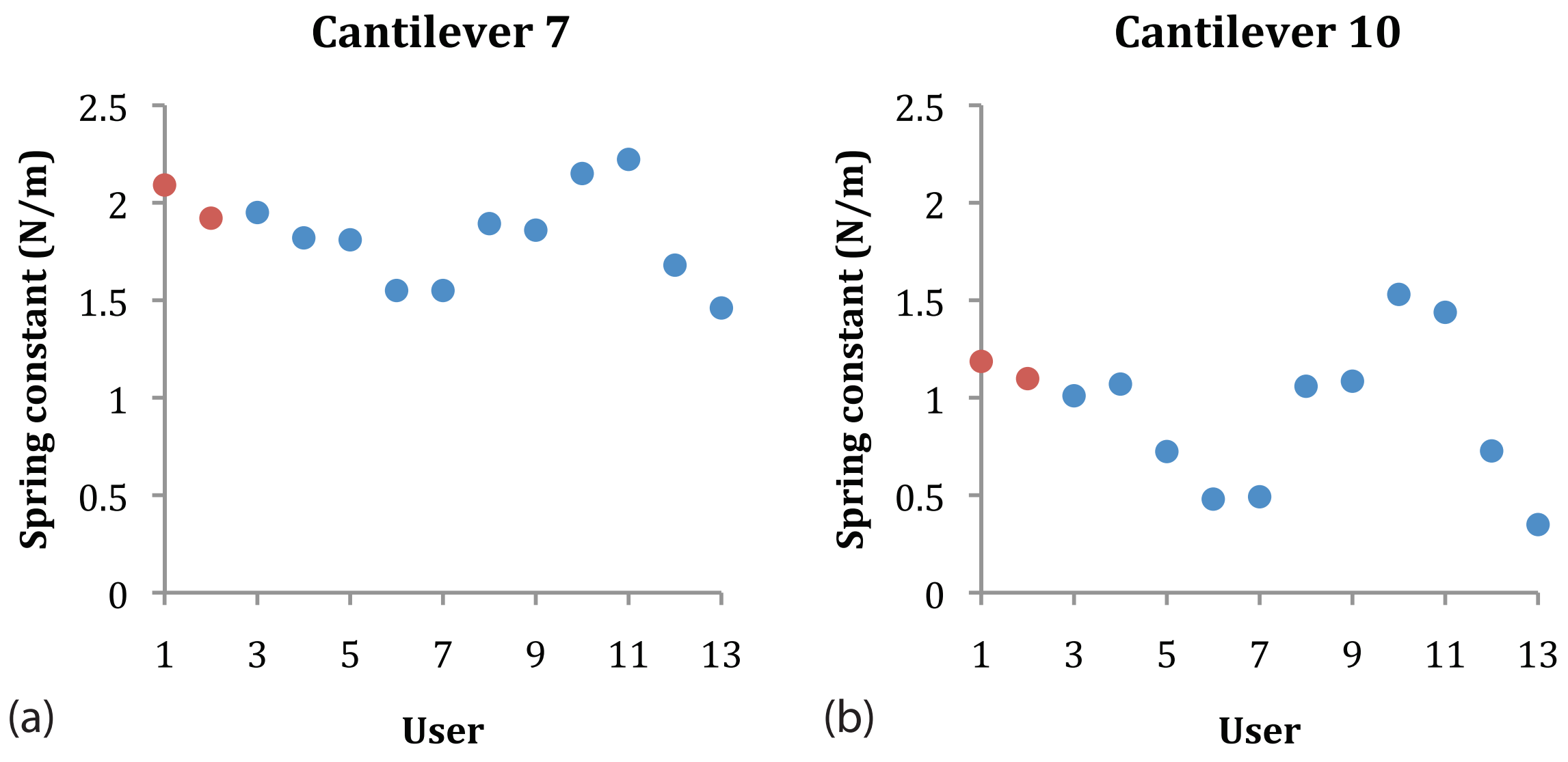

**FIG. 7:** Measured spring constant at the imaging tip position for two of the cantilevers using the LDV (Users 1 & 2 – red) and AFM Thermal method (blue). Six Groups and 13 Users performed the measurements. Spring constant of each cantilever is expected to be constant as a function of User number – barring small variations due to statistical uncertainty and temperature variations – because all Users calibrated exactly the same cantilever. Identical vertical scales are used for comparison.

These results show that the accuracy and precision of the AFM Thermal method can vary significantly between Users and Groups. This is most evident in the results of Cantilever 10, where the maximum and minimum measured spring constants differ by more than a factor of 4 (minimum is 0.35 N/m, maximum is 1.5 N/m – LDV measurement is 1.1 N/m); other cantilevers exhibit a maximum-to-minimum ratio of about 2, see Supplementary Information. The average relative SD in the measured spring constants of each cantilever using the AFM Thermal method, across all Users, is 19%. Cantilever 10 exhibits a higher relative SD of 40% across Users, and removing it from the total average gives a slightly lower value of 17%. Such error estimates are often used when analysing AFM force measurements, but clearly do not specify the accuracy of measurements from individual laboratories; see Fig. 7.

We remind the reader that each User calibrated exactly the same cantilever and had no access to data from other Users or Groups. The reported variation in Fig. 7 is therefore indicative of what can occur in the general AFM community, since users typically do not have (i) an independent benchmark by which to compare their calibration data, nor (ii) access to repeated measurements by other groups on the same cantilever. A previous round-robin study [5] also demonstrated that the accuracy of an individual AFM Thermal method measurement is instrument and user dependent – variations in the measured spring constant up to a factor of about two were reported between laboratories, although the differences were typically smaller. The reason for this strong variation could not be identified, though issues related to the $z$-piezo operation were suspected.

Spring constant measurements using the AFM Thermal method intrinsically fold in uncertainty due to (i) variation in calibration of the $z$-piezo scanner, and (ii) issues associated with using this scanner to calibrate the photodiode deflection detector; in addition to other complicating factors mentioned in Section I. It is therefore not surprising that spring constants measured using the AFM Thermal method, across the five groups in Fig. 7, fluctuate much more than the resonant frequencies and quality factors in Figs. 5 and 6, respectively.

This highlights the practical utility of the GCI where users can (in real-time) standardise their spring constant measurements to a live value derived from independent groups across the AFM community. It also emphasises the intrinsic advantage of a calibration method that relies on the measured resonant frequency and quality factor only.

*d. A-coefficient* Figure 8 gives results for the (universal) $A$-coefficient derived from the measurements in Figs. 5 – 7, using Eq. (3). Despite the LDV-measured spring constants of cantilevers 7 and 10 differing by a factor of 2, the $A$-coefficients in Fig. 8 are found to exhibit a mean close to that determined by LDV, i.e., $A = 6.4\ \mathrm{nN\ s^{1.3} m^{-1}}$ [see Eq. (4)]. The considerable scatter of data in Fig. 8 is predominantly due to uncertainty in the spring constant measured by the AFM Thermal method; see above. Nonetheless, the results in Fig. 8 show that with appropriate averaging, as per Eq. (3), such fluctuations in the $A$-coefficient are suppressed.

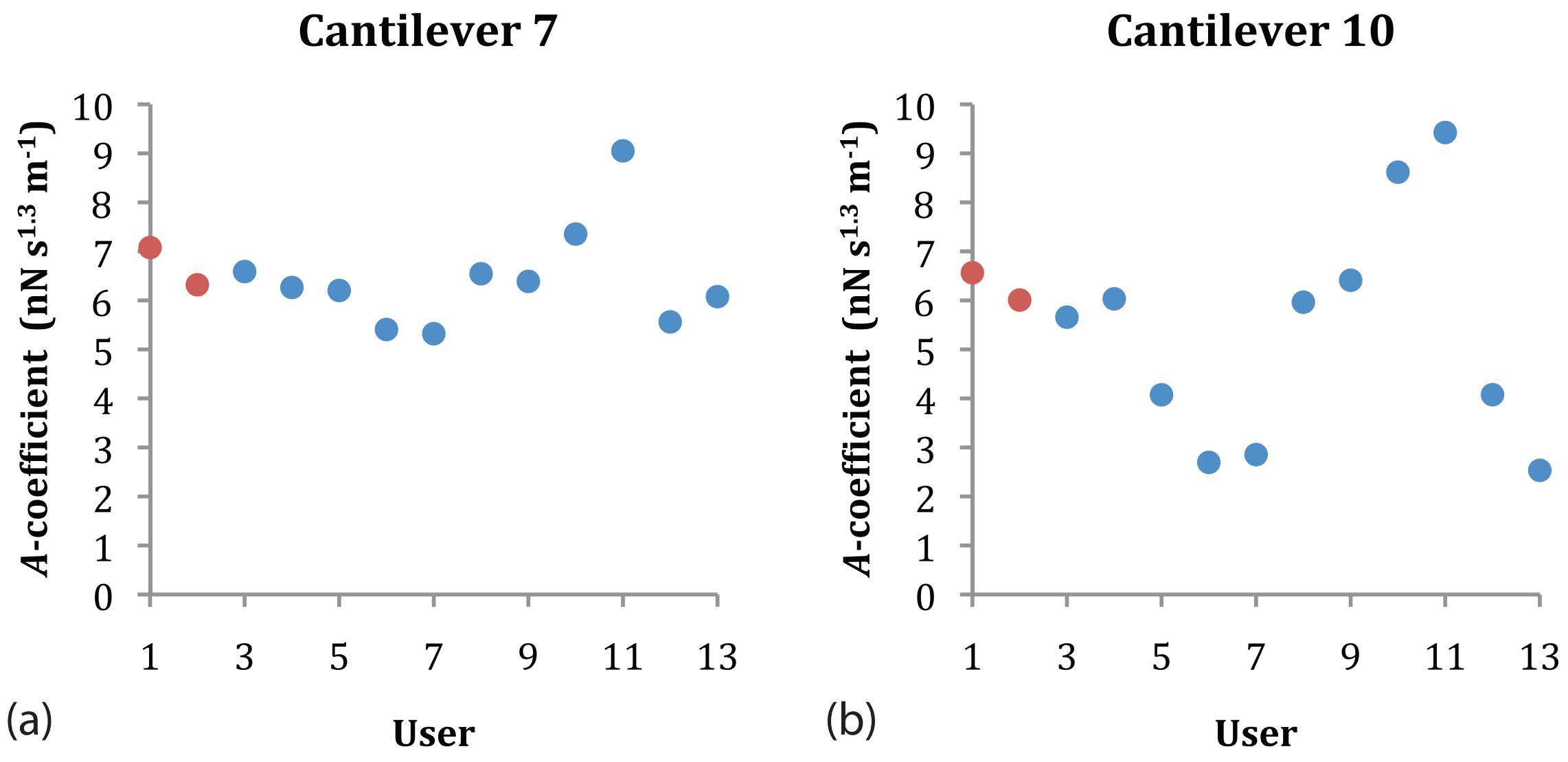

**FIG. 8:** Measured $A$-coefficient for two of the cantilevers using the LDV (Users 1 & 2 – red) and AFM Thermal method (blue). Six Groups and 13 Users performed the measurements. Identical vertical scales are used for comparison.

Figure 9(a) presents a histogram of the measured $A$-coefficients of all cantilevers, as determined by the 5 Groups using the AFM Thermal method (LDV measurements are not included). Averaging this dataset, as per Eq. (3), gives

$$A = 6.38 \pm 0.27 \; [\mathrm{nN\ s^{1.3} m^{-1}}] \,, \tag{5}$$

with the listed uncertainty again specifying a 95% confidence interval. This is in excellent agreement with the (independent) LDV measurement reported in Eq. (4).

While the LDV measurement (of the same cantilevers) uses fewer data-points (20 in total, in contrast to 110 in the AFM Thermal method), the overall uncertainty in its $A$-coefficient is lower. This is because the LDV typically produces a spring constant of superior accuracy and precision. Indeed, the LDV data in Fig. 4 sits in a narrow band on Fig. 9(a) spanning only the three central columns ($5.5 < A < 7$ nN $\mathrm{s^{1.3} m^{-1}}$). Importantly, the excellent agreement between Eq. (4) [LDV] and Eq. (5) [AFM] demonstrates the validity of using the AFM Thermal method and multiple cantilevers to determine the $A$-coefficient.

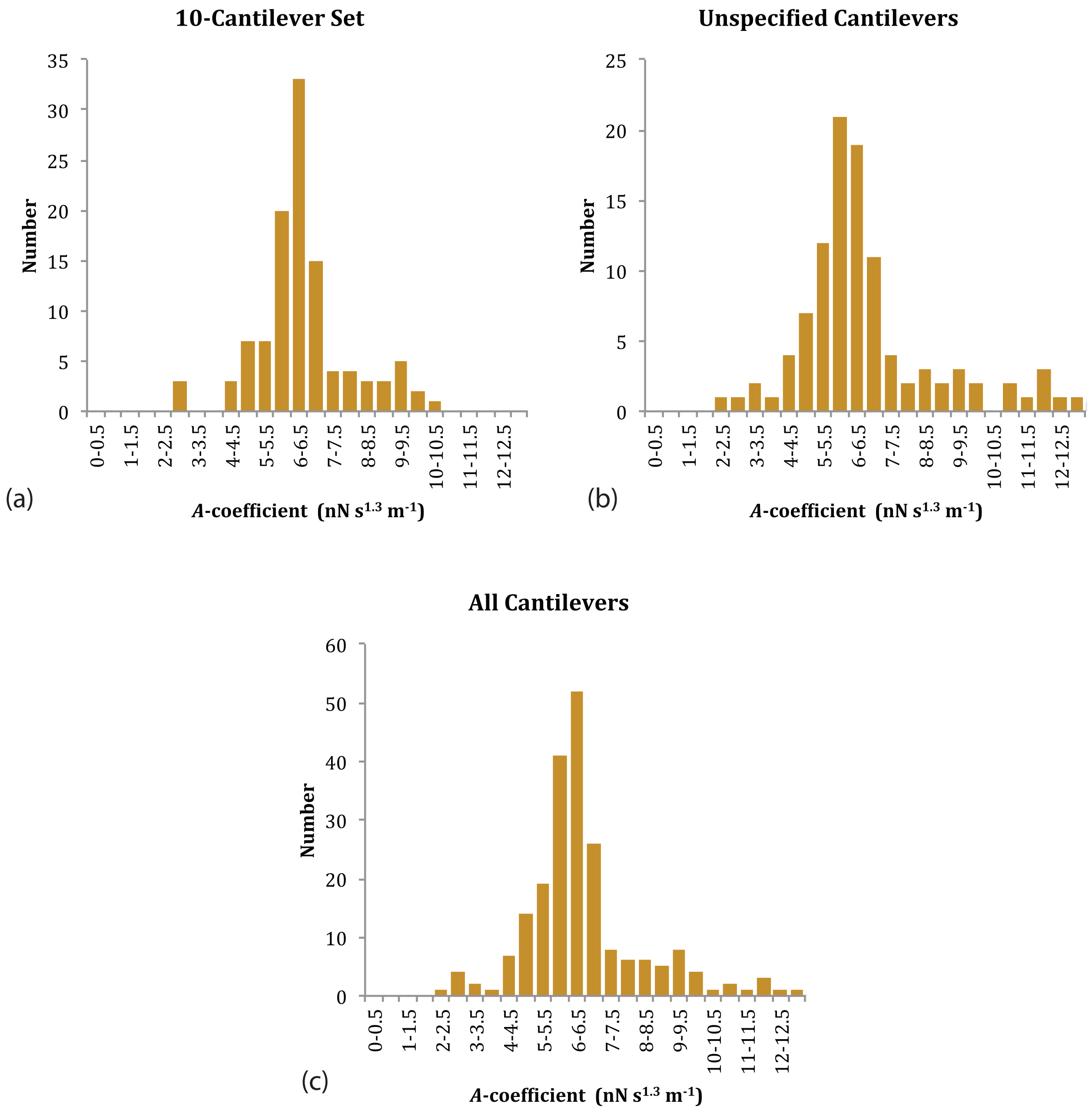

**FIG. 9:** Histogram of measured $A$-coefficients for (a) 10-cantilever set, (b) unspecified cantilevers, and (c) all cantilevers. Measurements performed using the AFM Thermal method. (a) 5 Groups and 11 Users giving 110 data-points, (b) 4 Groups and 8 Users giving 103 data-points, and (c) measurements in previous two sets collated giving 213 data-points. Horizontal scales are identical in all cases.

## B. Unspecified AC240-R3 cantilevers

Equations (2) and (3) are applicable to cantilevers of the same type that do not have identical plan view dimensions, provided the dimensional variations are not large; this is discussed in Ref. [11]. Such variations are expected between different, yet identical type cantilevers, across the AFM community. Using a range of randomly procured cantilevers of the same type is expected to increase scatter in the measured $A$-coefficients, which can be reduced through averaging. This expectation is examined in this section.

Groups 2 – 5 independently procured additional AC240-R3 cantilevers and calibrated each of those cantilevers using the AFM Thermal method, yielding 103 data-points for the $A$-coefficient; see histogram in Fig. 9(b). A total of 8 separate Users performed the measurements. Averaging this dataset as per Eq. (3) gives

$$A = 6.50 \pm 0.39 \; [\mathrm{nN\,s^{1.3} m^{-1}}] , \tag{5}$$

where the uncertainty again specifies a 95% confidence interval. While the standard error is significantly larger than the AFM Thermal method measurements of the 10-cantilever set (for a similar sample size of ~100 data-points), the agreement between Eqs. (4) and (5) is striking and is well within the determined uncertainty. Table II provides a summary of the $A$-coefficients measured using all approaches.

**TABLE II:** $A$-coefficients measured using the different approaches. The specified uncertainty gives a 95% confidence interval. The "LDV of 10 cantilevers" result is derived from $N = 20$ data-points, "AFM of 10 cantilevers" from $N = 110$ data-points, "AFM of unspecified cantilevers" from 103 data-points, and "AFM of all cantilevers" from $N = 213$ data-points.

| Approach | $A$-coefficient (nN $s^{1.3}$ $m^{-1}$) | $N$ |
|---|---|---|
| LDV of 10-cantilever set | $6.43 \pm 0.15$ | 20 |
| AFM of 10-cantilever set | $6.38 \pm 0.27$ | 110 |
| AFM of unspecified cantilevers | $6.50 \pm 0.39$ | 103 |
| AFM of all cantilevers | $6.44 \pm 0.23$ | 213 |

Finally, we combine the data reported in Figs. 9(a) and 9(b), which are obtained using the AFM Thermal method for the 10-cantilever set and the unspecified cantilevers, respectively, to produce a histogram of all AFM data in Fig. 9(c). Averaging this collective data, which excludes the LDV measurements, gives

$$A = 6.44 \pm 0.23 \; [\mathrm{nN\,s^{1.3} m^{-1}}] , \tag{6}$$

which also agrees well with the LDV measurement of the 10-cantilever set in Eq. (4); see Table II.

This analysis demonstrates that unspecified cantilevers of a given type (plan view geometry) can be used to accurately determine the $A$-coefficient. This finding is important because the GCI must generally use unspecified cantilevers to determine the $A$-coefficient from independent users across the AFM community.

Values for the $A$-coefficient in Table II are slightly larger (by ~10%) than those reported in Ref. [13] that were based on measurements of four AC240-R3 cantilevers only. Note that Eq. (3) is weakly dependent on the cantilever's plan view dimensions, as discussed in Section II. The plan view dimensions of the cantilevers used in Ref. [13] were not measured and a rigorous assessment of the reason for this difference cannot therefore be made. This difference highlights the importance of sampling a large number of cantilevers, as required in the GCI, to eliminate statistically insignificant fluctuations that inevitably occur when a small number of data-points are used to determine the $A$-coefficient. The GCI inherently generates and uses an average $A$-coefficient because small variations in geometry and dimensions of a particular cantilever always occur.

## V. SUMMARY AND CONCLUSION

Spring constant calibration of cantilevers is often conducted without reference to measurements from other laboratories or a global standard. The round-robin study presented here, using a single set of 10 cantilevers, highlights the variability in current AFM force measurements that can hinder robust comparison between independent groups and users.

The Global Calibration Initiative described in this article allows all users in the AFM community to (i) compare their calibration results to those of others, thus standardising their AFM force measurements, and (ii) non-invasively calibrate any cantilever type. Proof-of-principle demonstration of the GCI is reported on a single cantilever type using independent measurements from five groups across three countries.

All users in the AFM community can propose other cantilever types for use in the GCI, enabling standardisation across available/future models and their non-invasive calibration. The URL for the GCI website is provided in Ref. [14].

## ACKNOWLEDGMENTS

Authors acknowledge support from the MicroNano Research Facility (MNRF), RMIT University, and Babs Fairchild; ADS, CJS, CTG, Heyou Z, Hongrui Z, JES, JT, MJH, PM, TZ from the Australian Research Council Grants Scheme; ADS, CJS, CTG from the Australian Microscopy and Microanalysis Research Facility; DBH, PAT, RB from the Knut and Alice Wallenberg Foundation; VR from the Olle Engqvist Foundation and the Swedish Research Council; JIK from Science Foundation Ireland (SFI12/IA/1449) and the Nanoscale Electrical Metrology project funded by the Collaborative Centre for Applied Nanotechnology (CCAN); MJH from an ARC Australian Research Fellowship, the Australian National Fabrication Facility and the ARC Centre of Excellence for Electromaterials Science (University of Wollongong).

# Supplementary Information for

# A virtual instrument to standardise the calibration of atomic force microscope cantilevers

John E. Sader, Riccardo Borgani, Christopher T. Gibson, David B. Haviland, Michael J. Higgins, Jason I. Kilpatrick, Jianing Lu, Paul Mulvaney, Cameron J. Shearer, Ashley D. Slattery, Per-Anders Thoren, Jim Tran, Heyou Zhang, Hongrui Zhang and Tian Zheng

## AFM Thermal method measurements of 10-cantilever set by all Users

### 1. Resonant frequencies (LDV: red; AFM: blue)

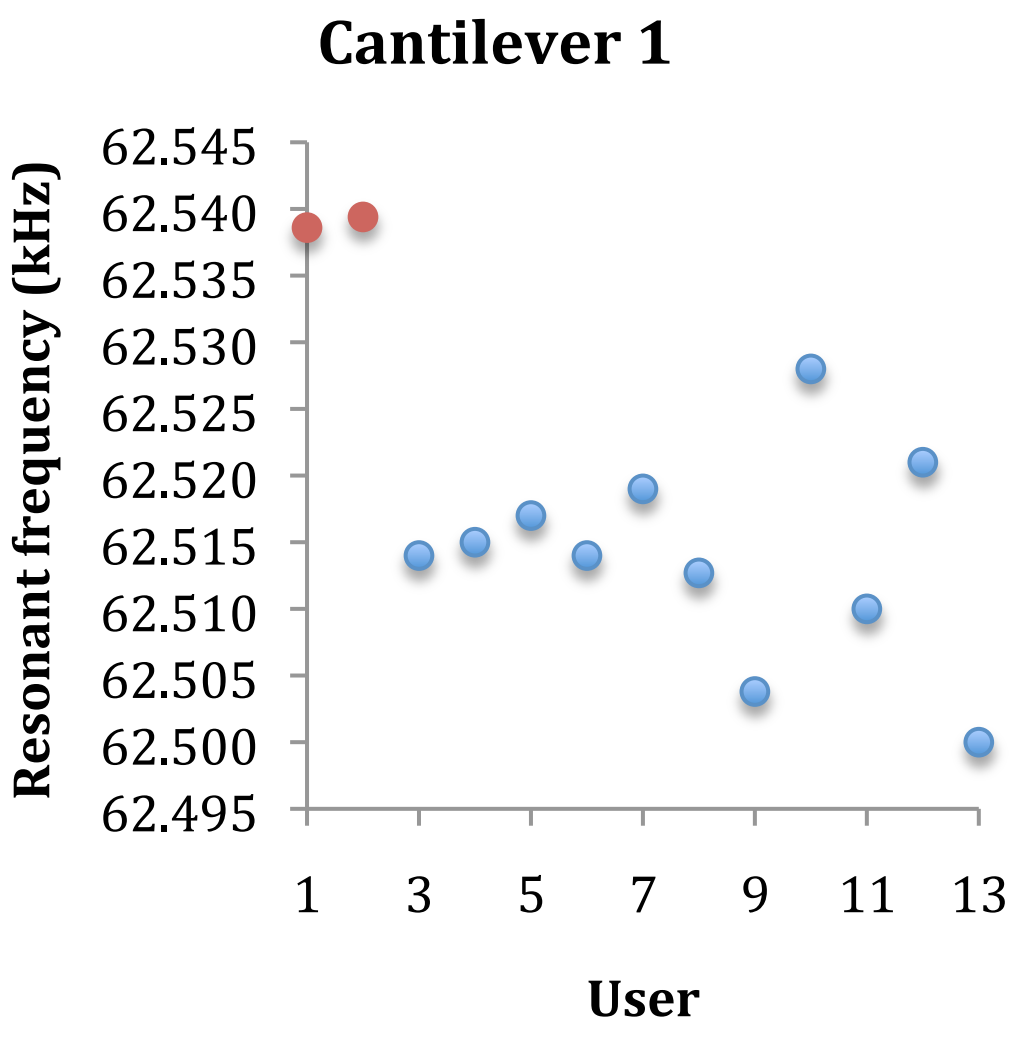

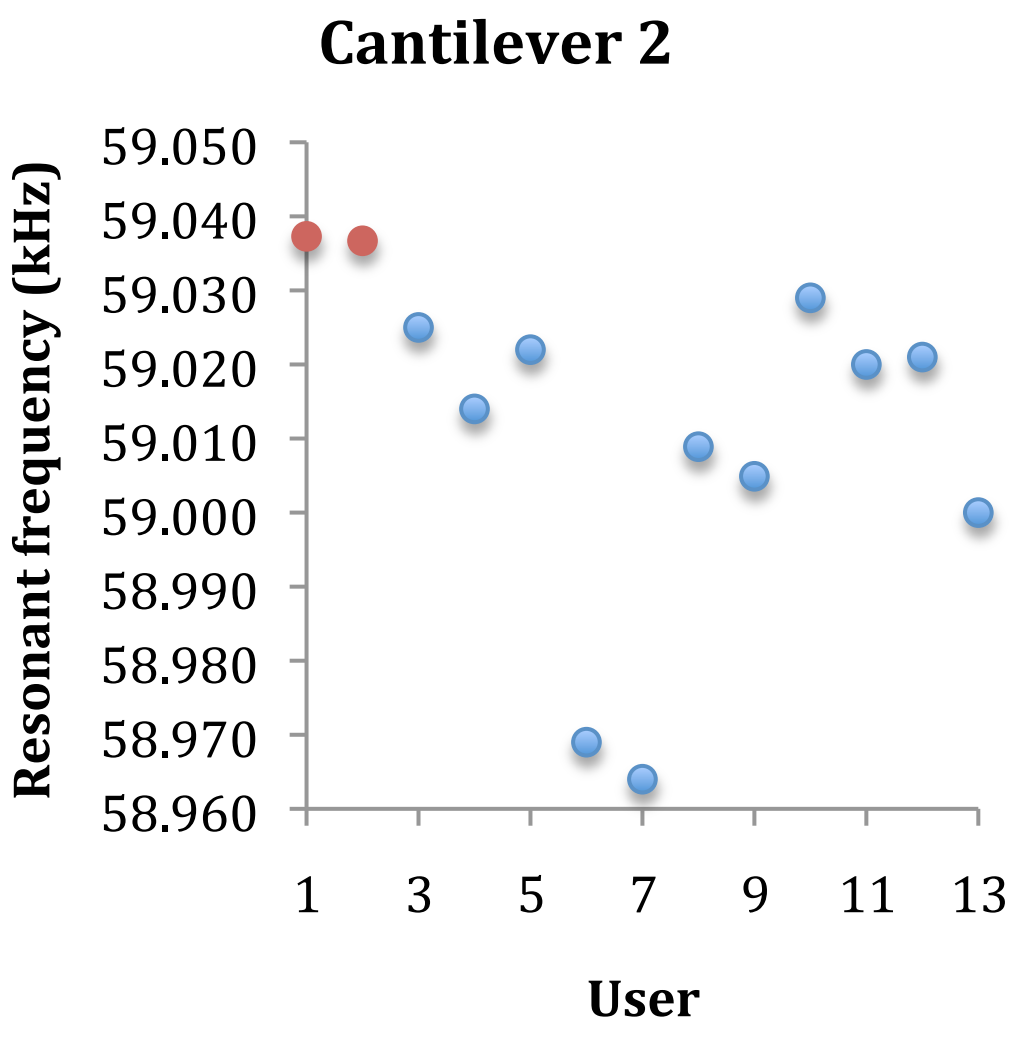

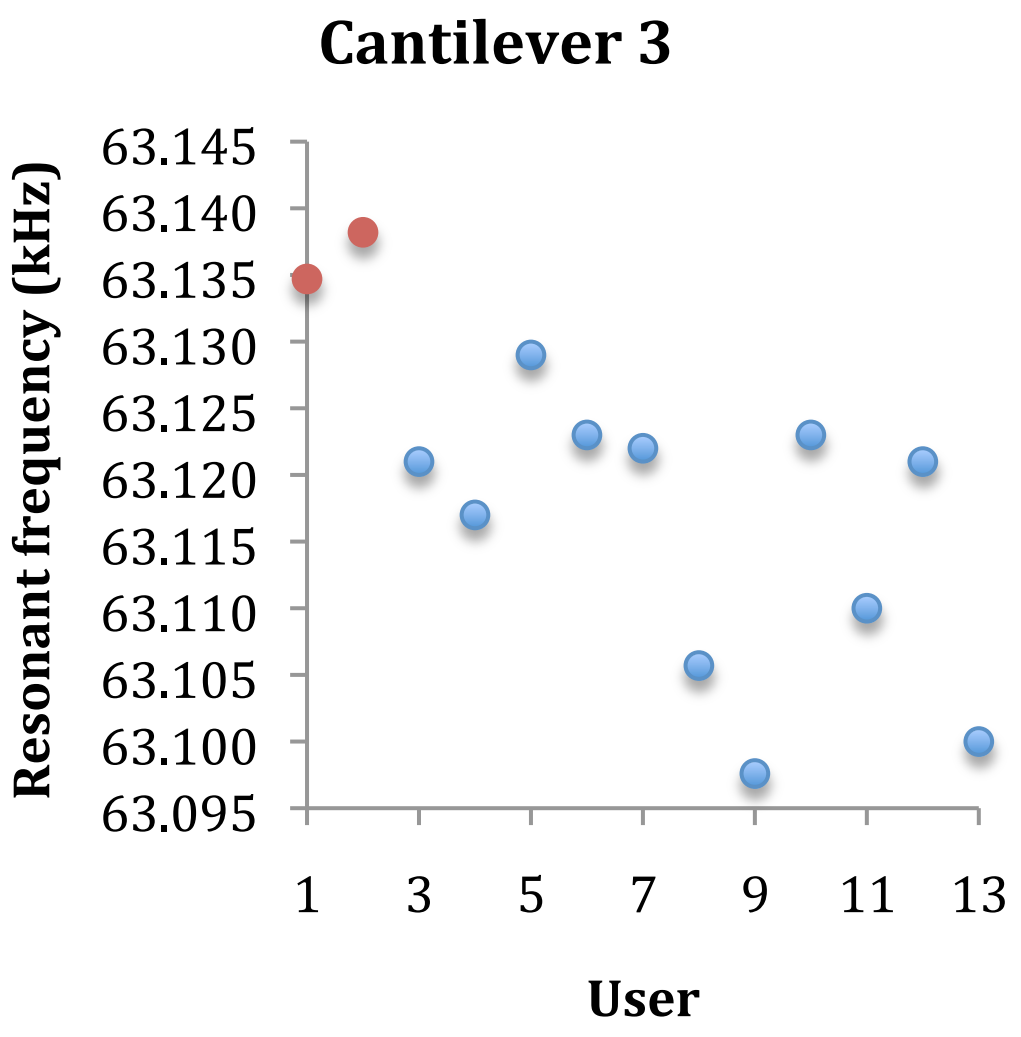

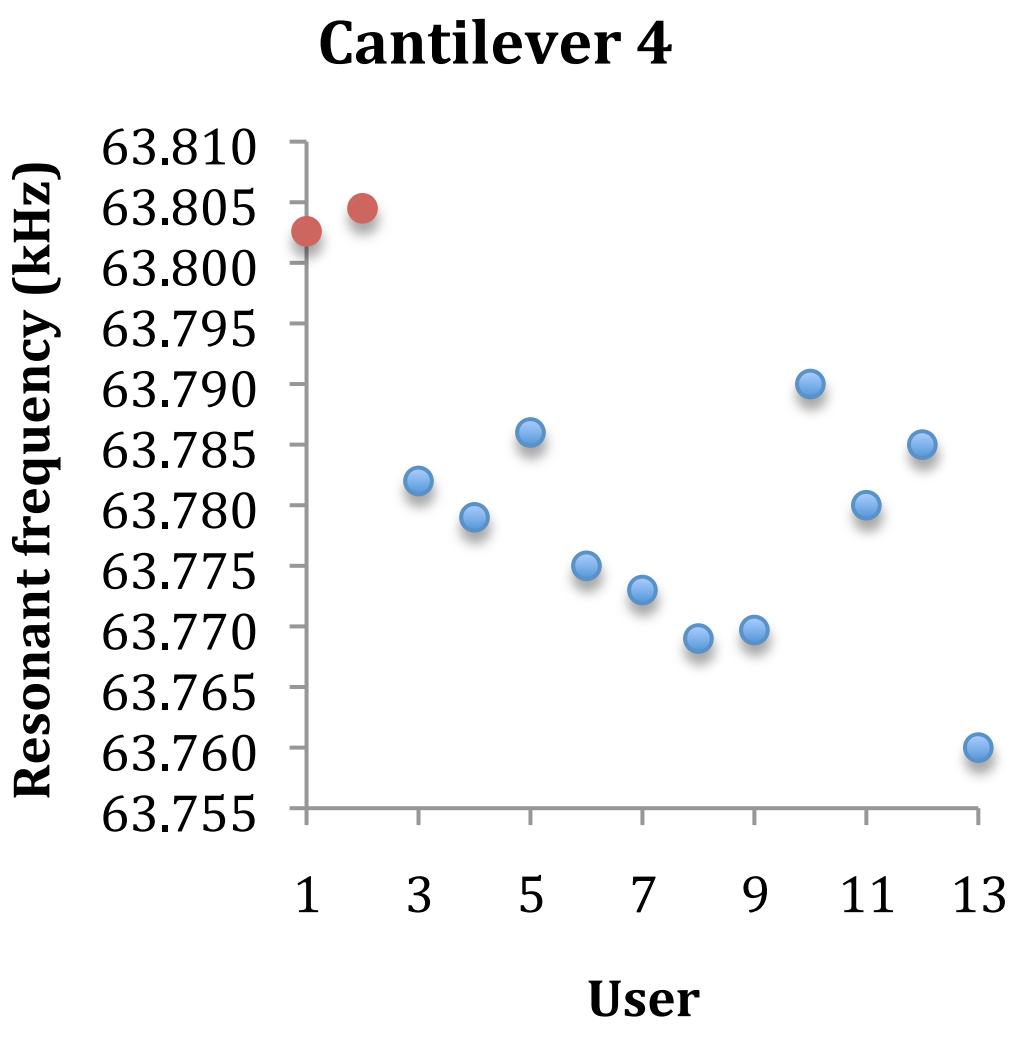

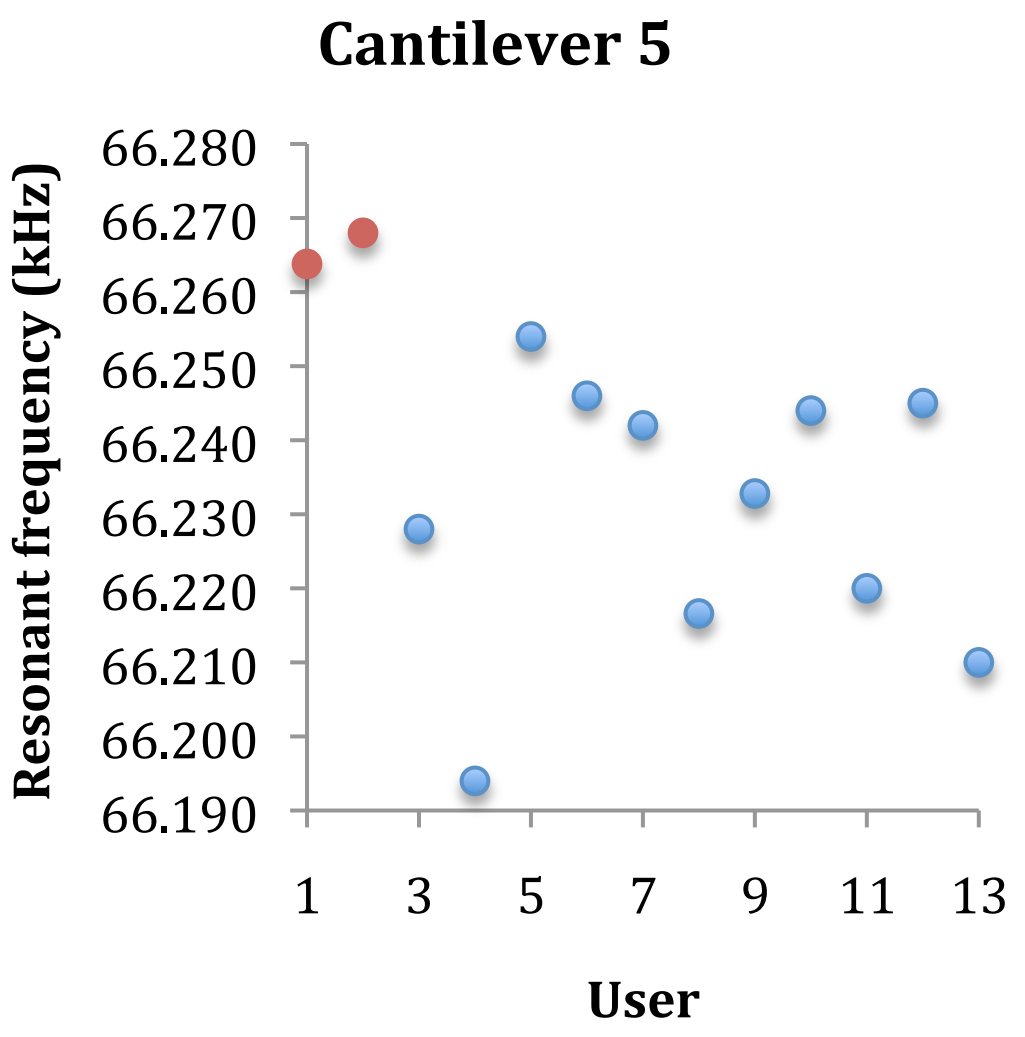

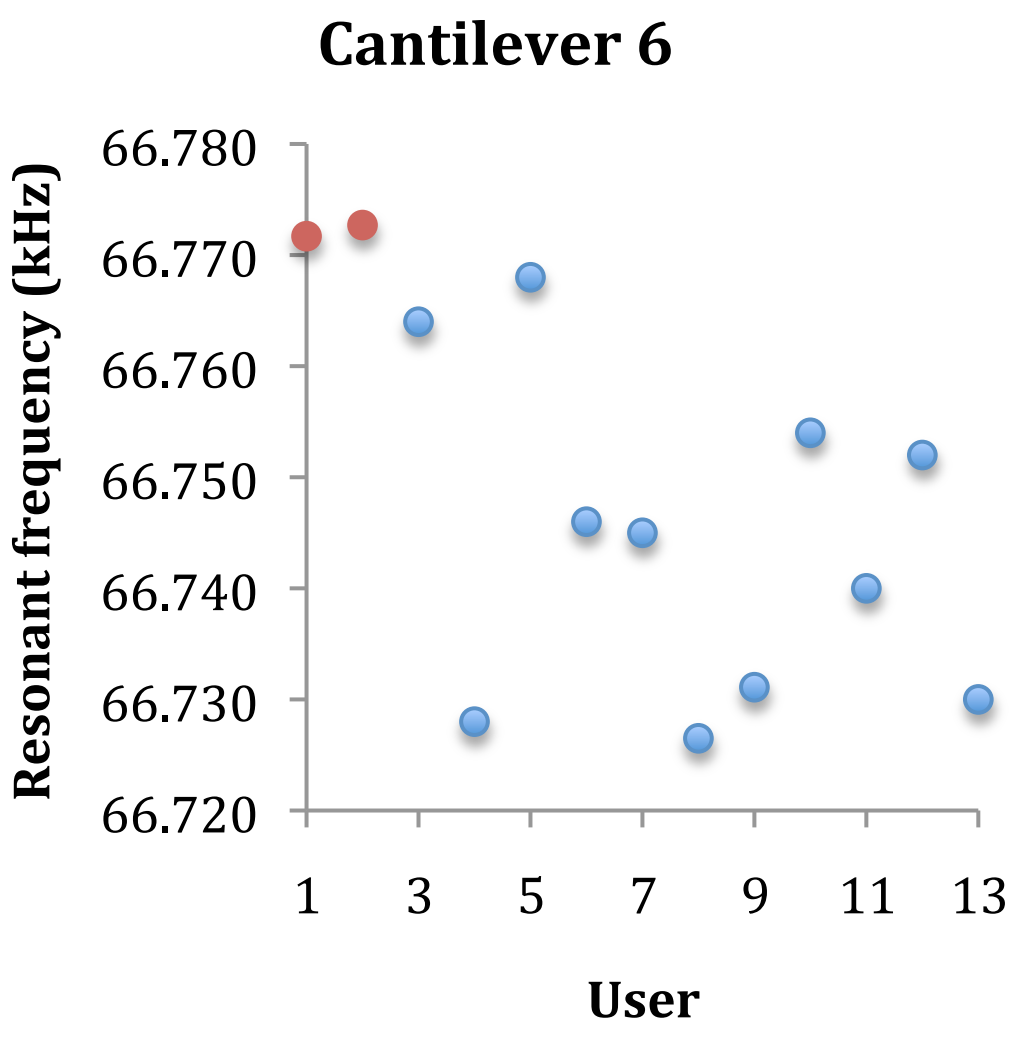

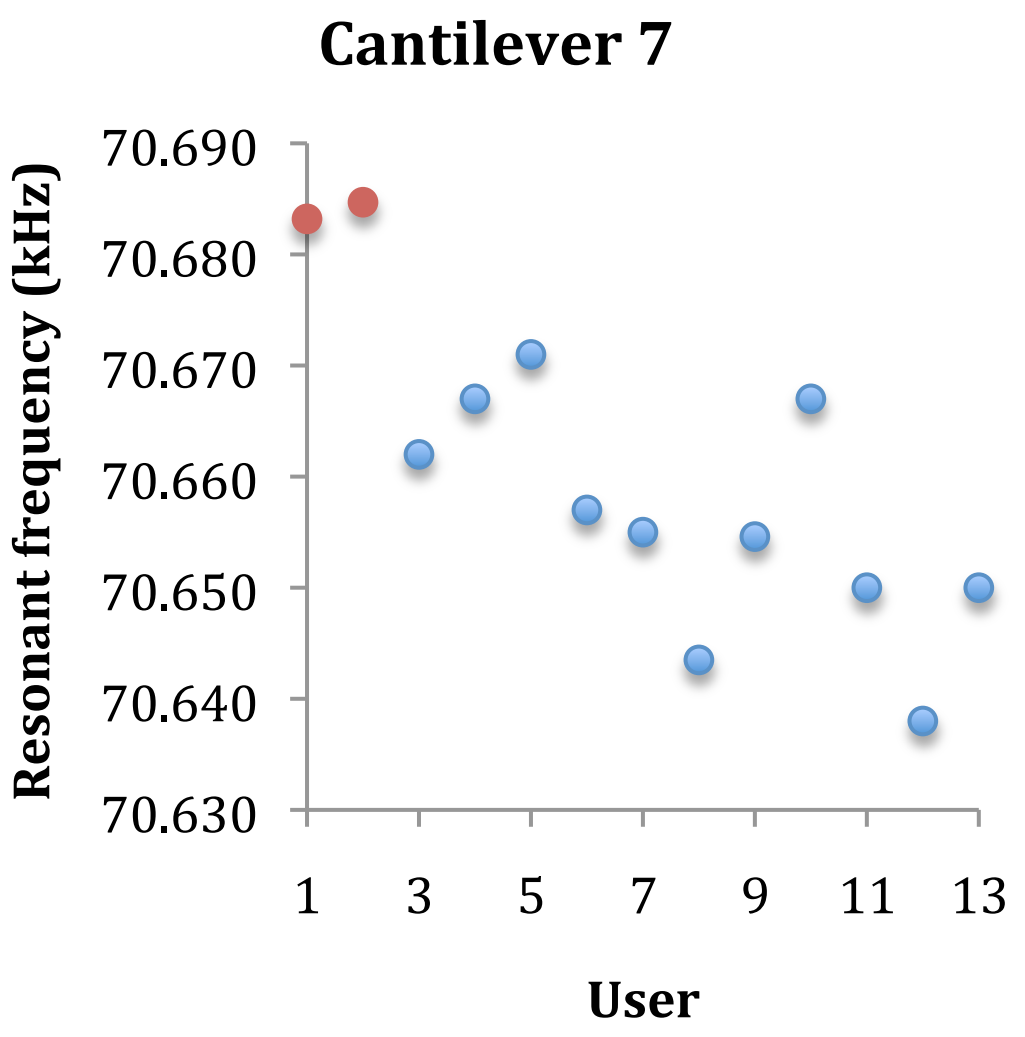

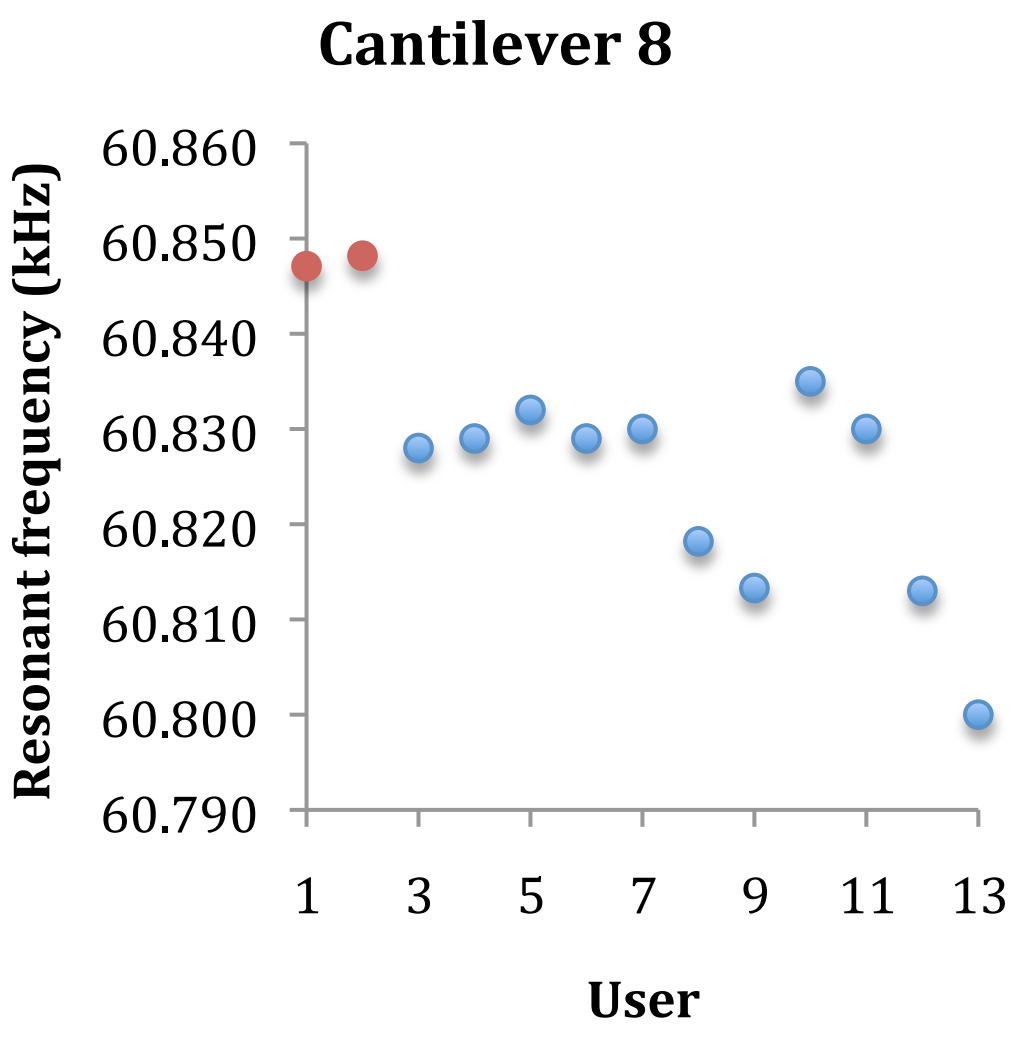

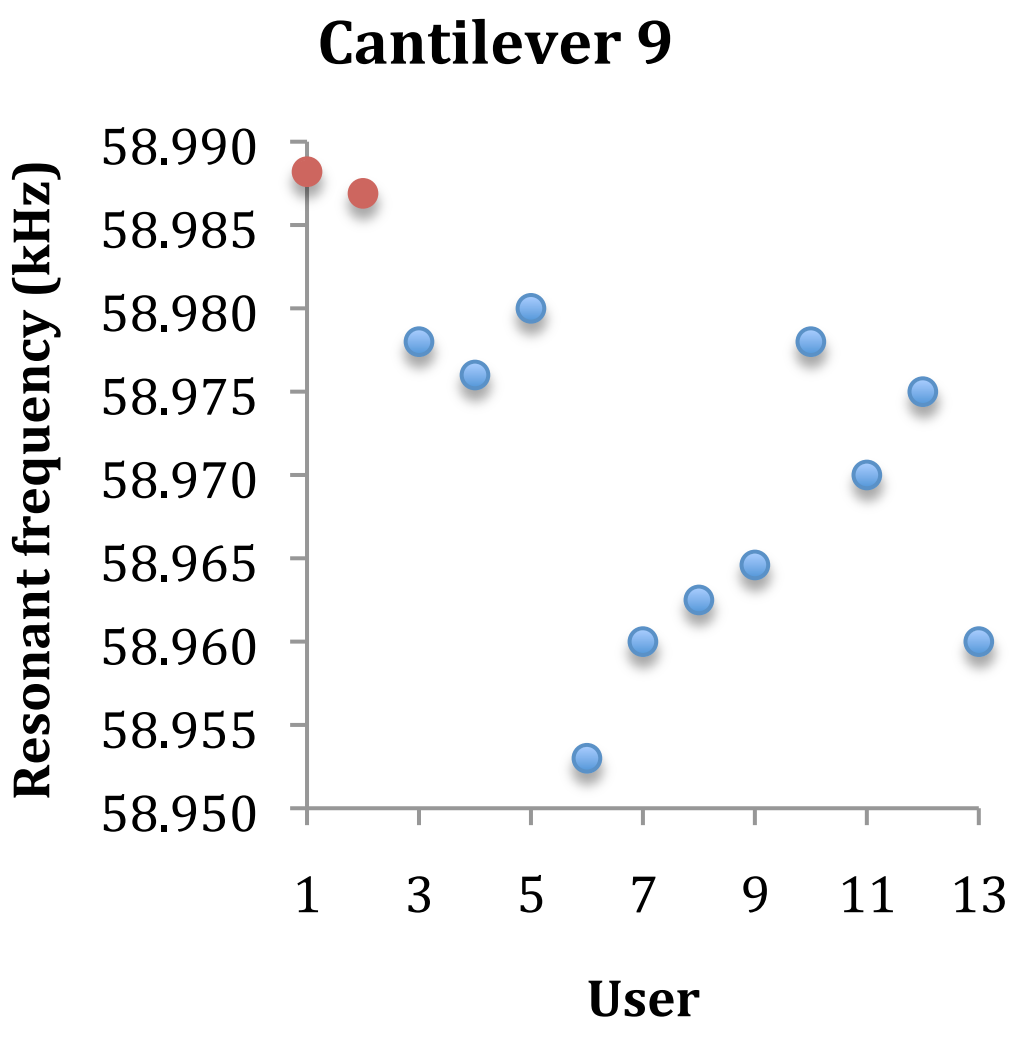

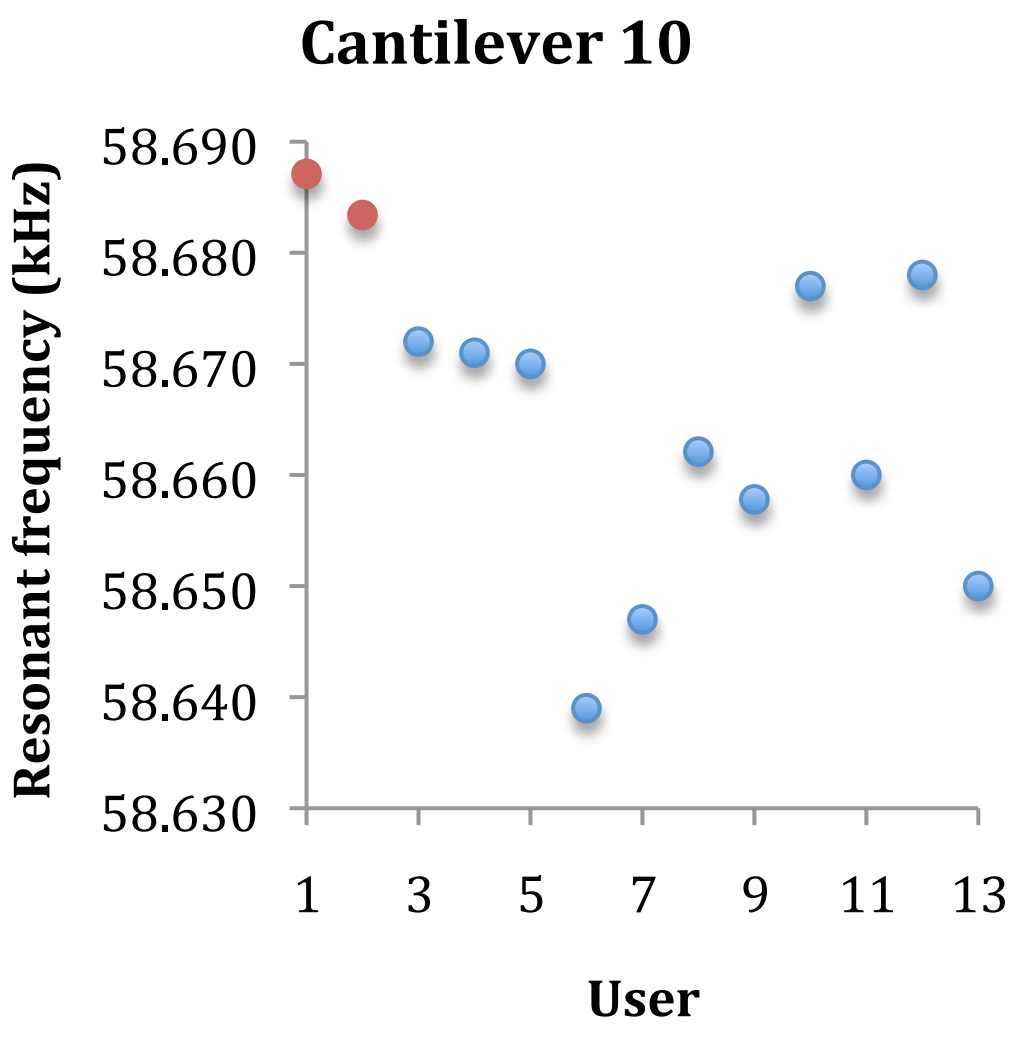

## 2. Quality factors (LDV: red; AFM: blue. Identical vertical scales used in all plots)

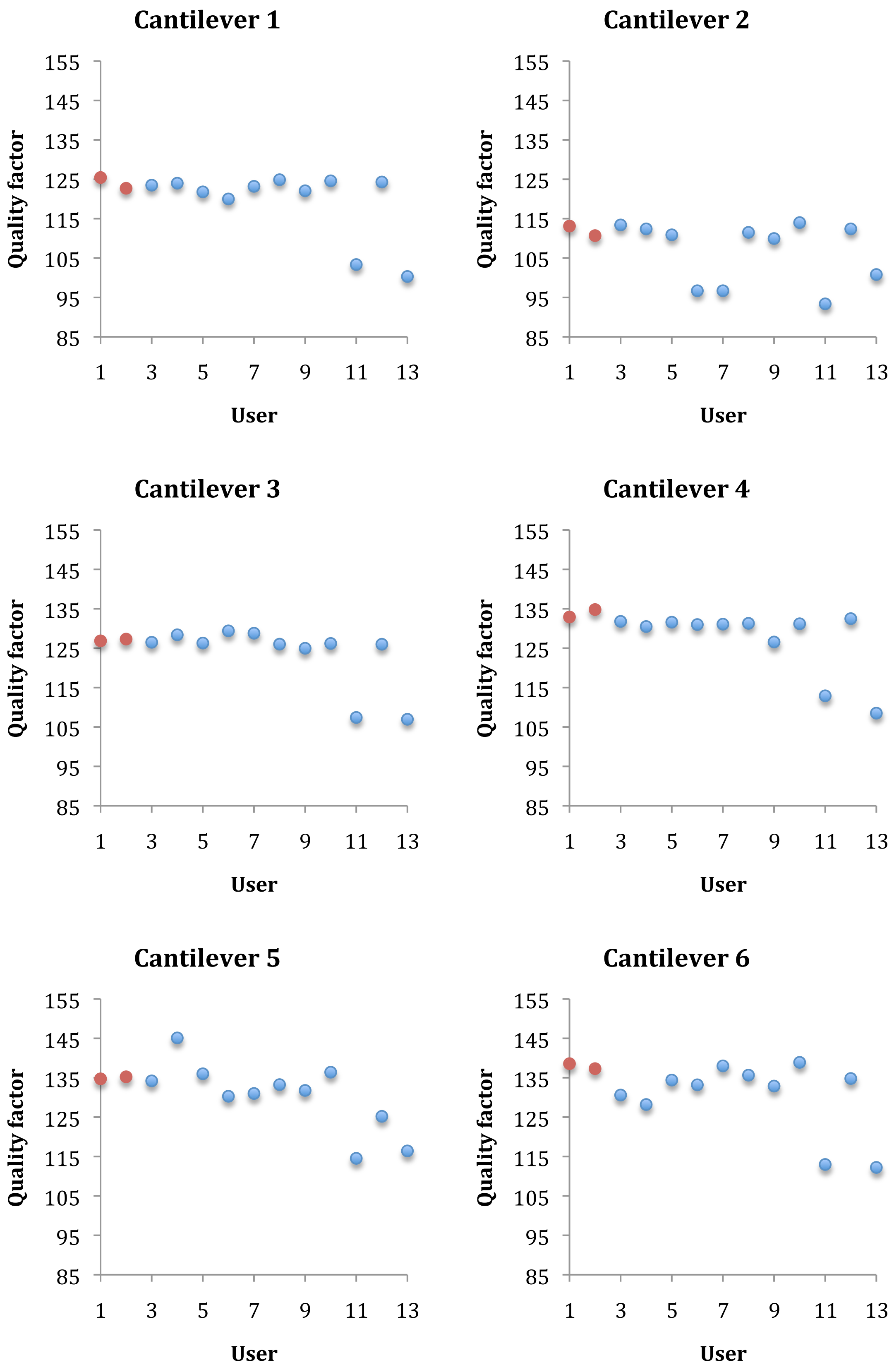

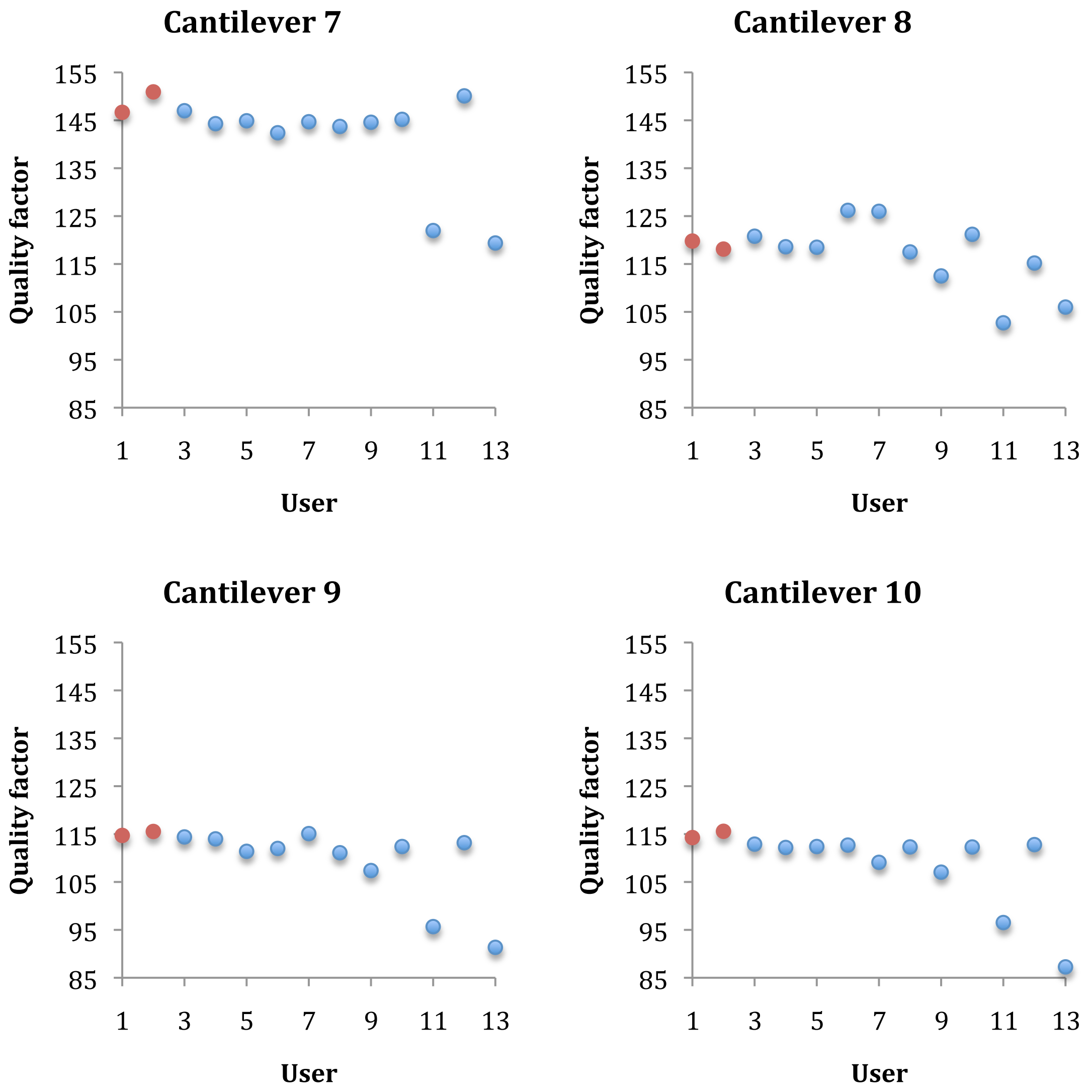

### 3. Spring constants (LDV: red; AFM: blue. Identical vertical scales used in all plots)

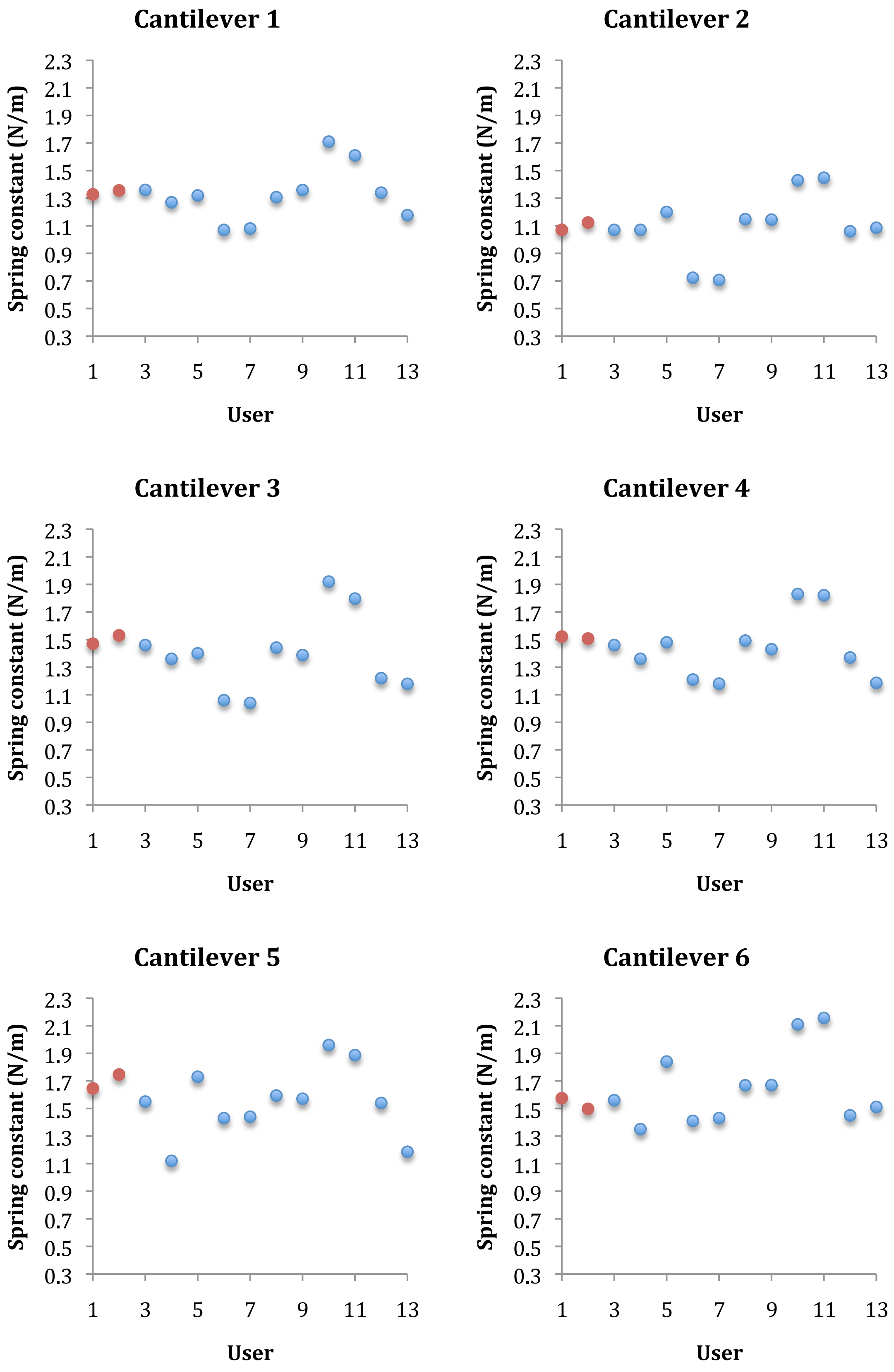

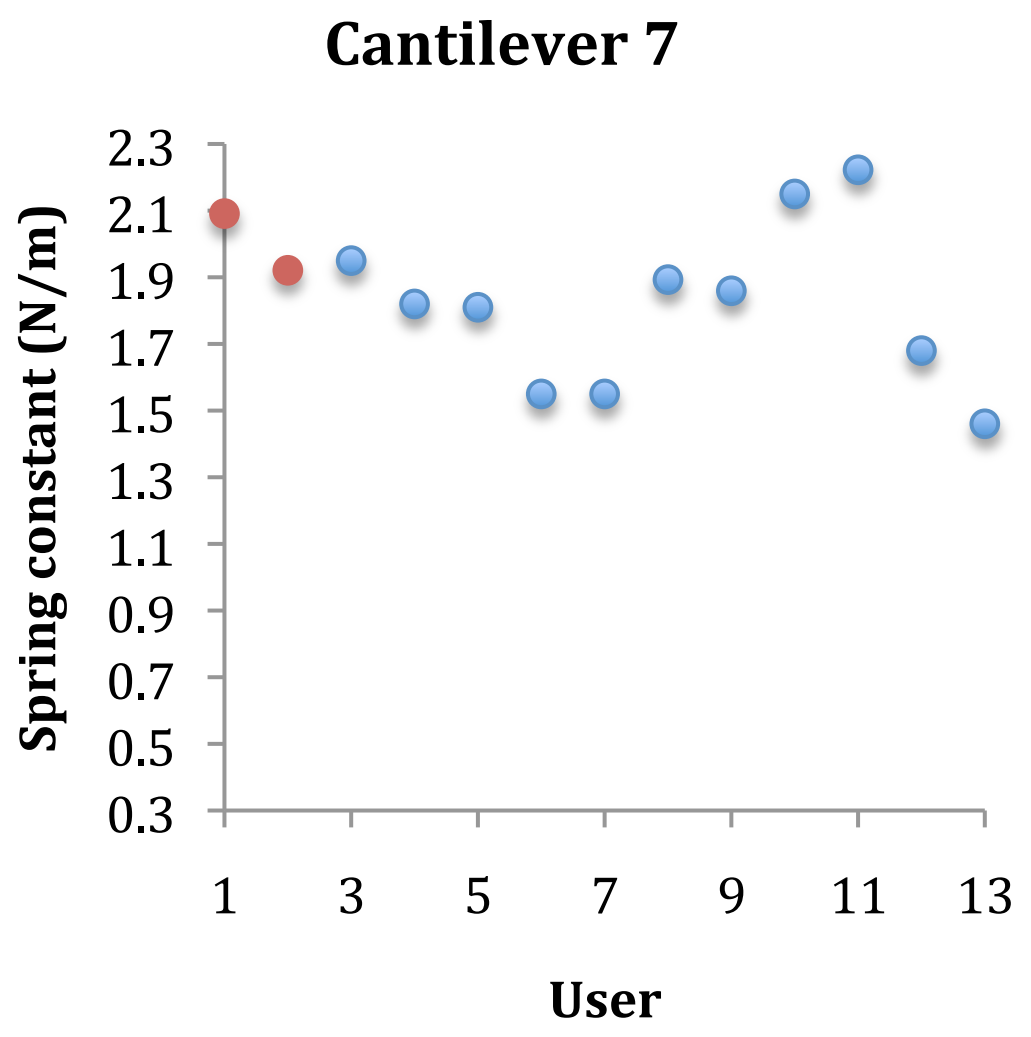

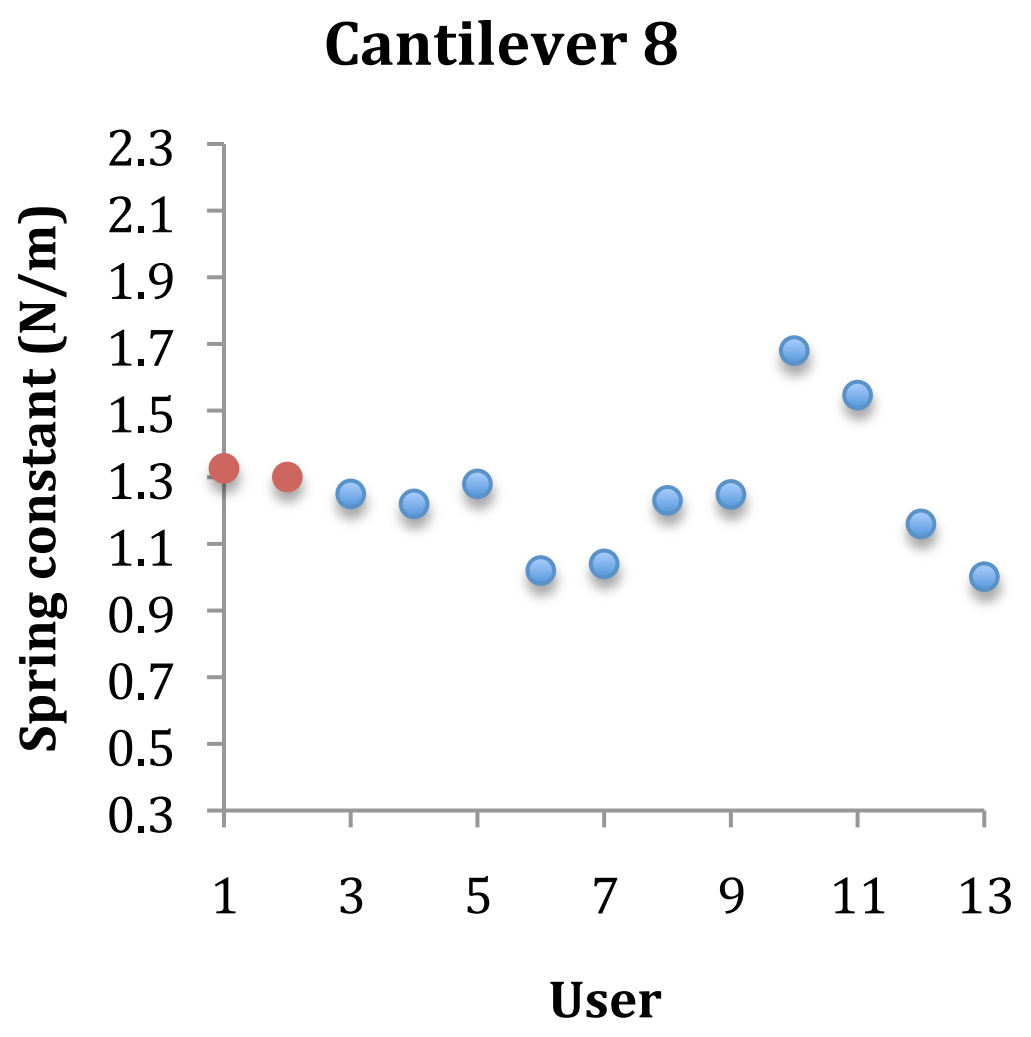

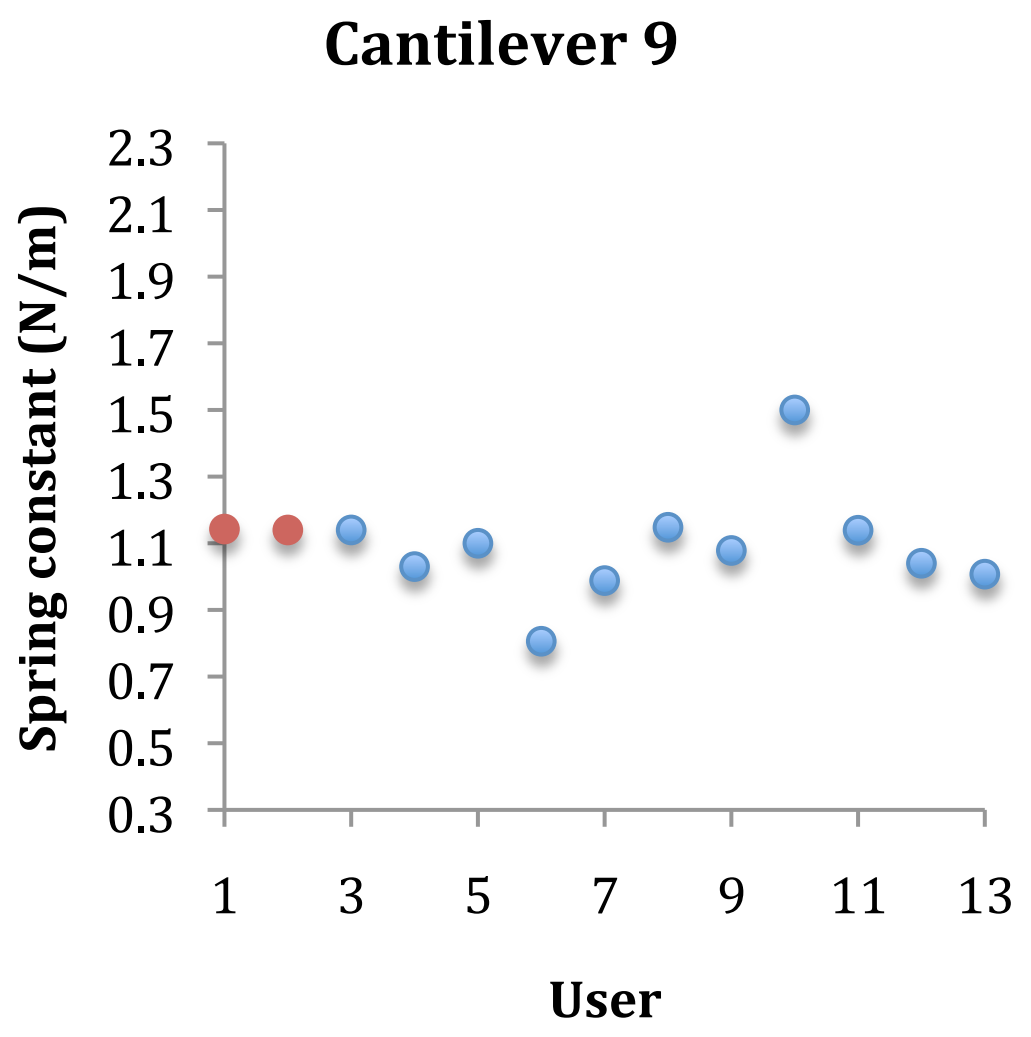

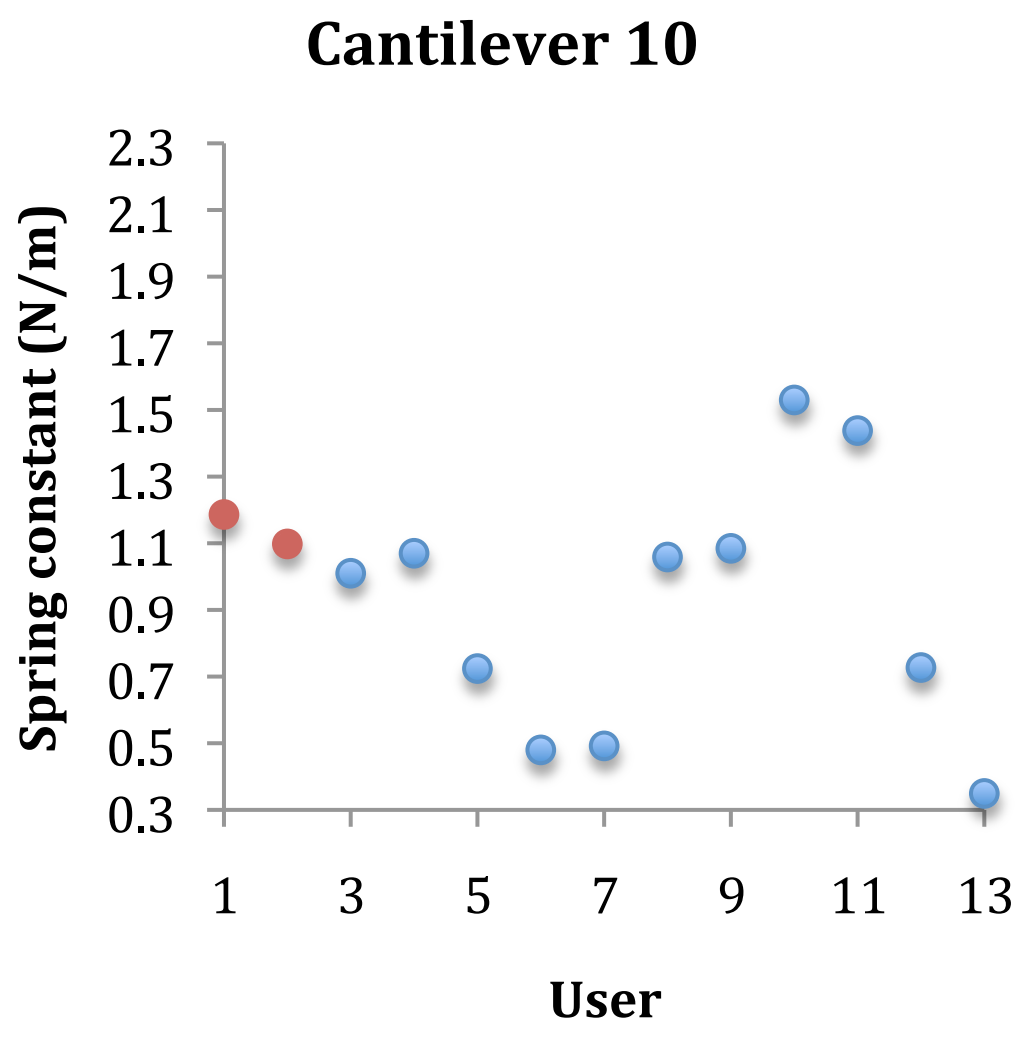

4. ***A*-coefficient (LDV: red; AFM: blue. Identical vertical scales used in all plots)**

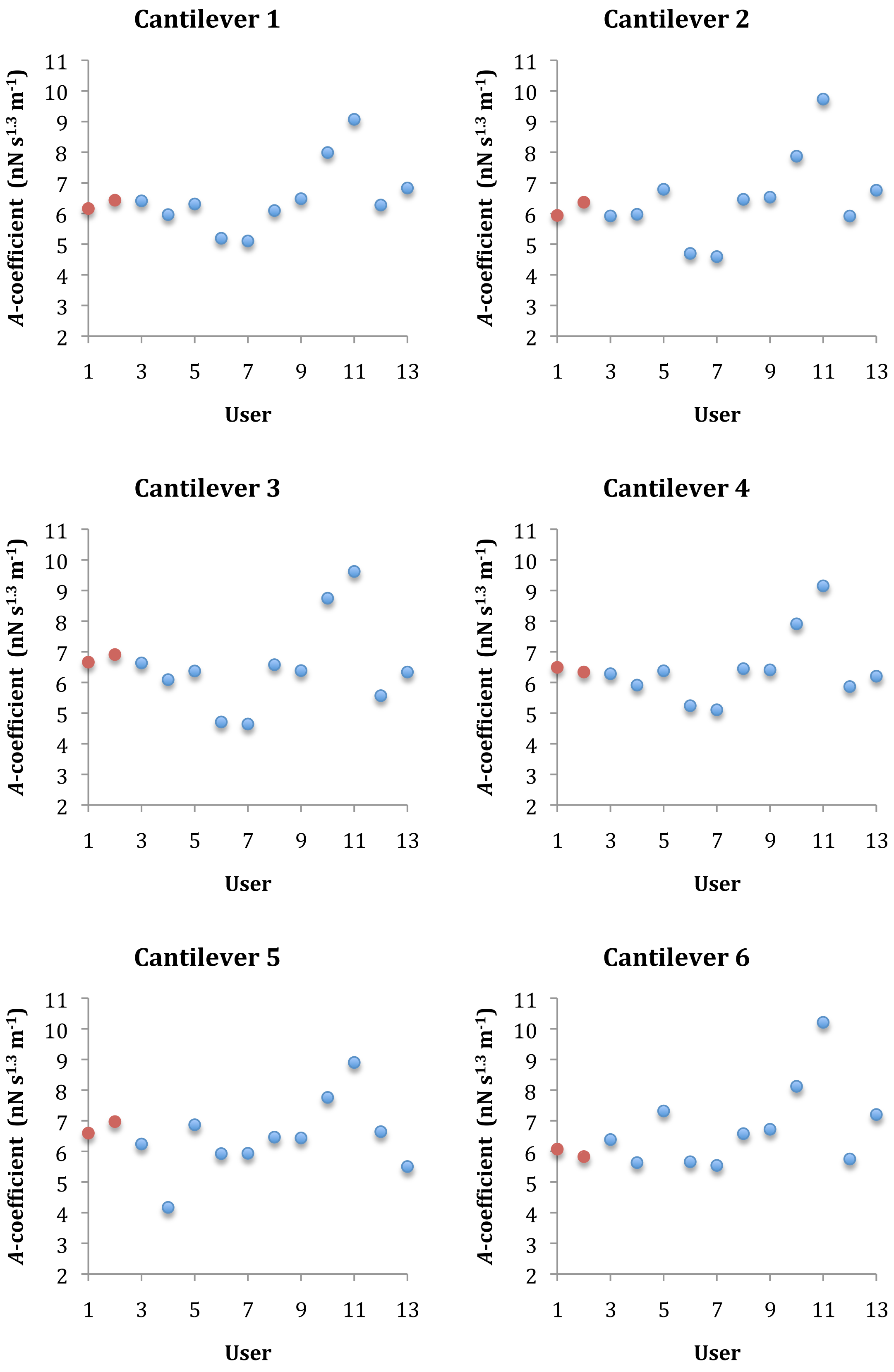

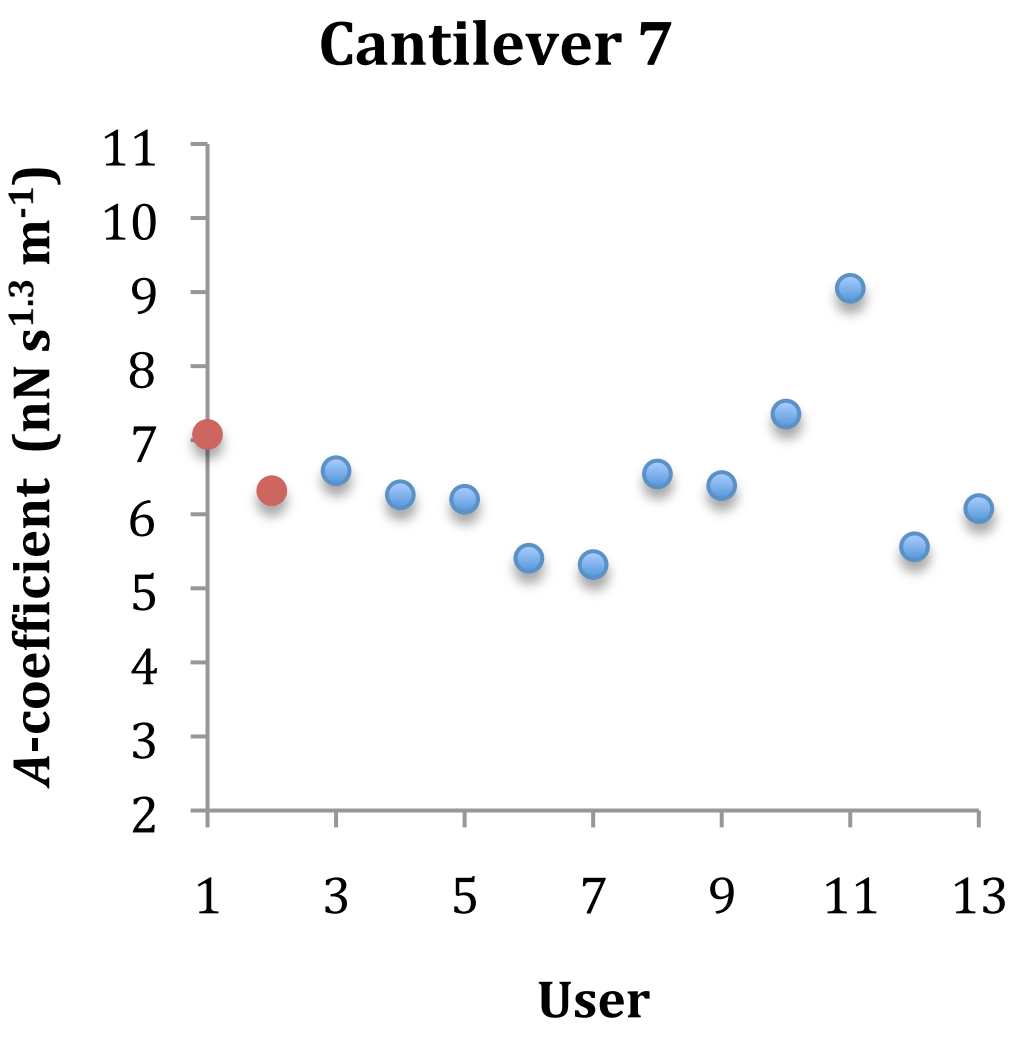

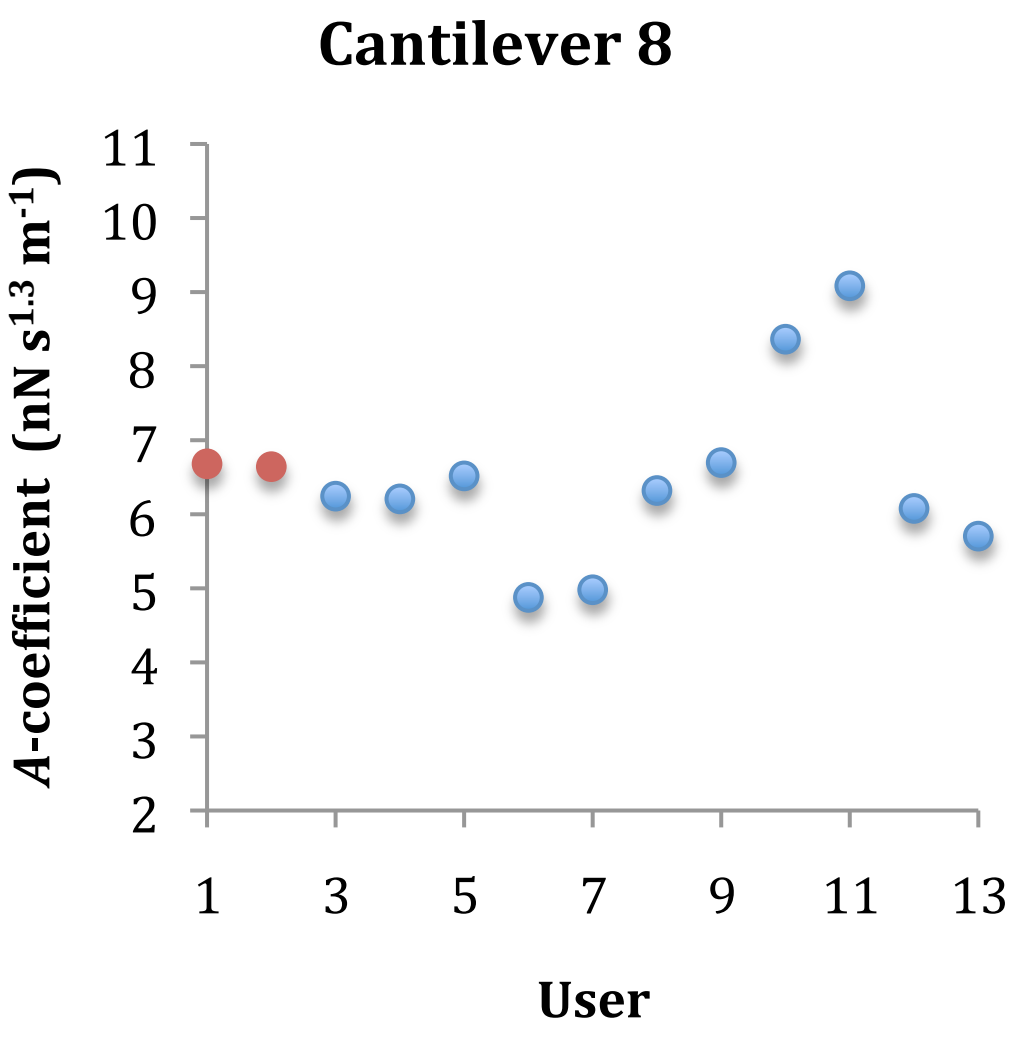

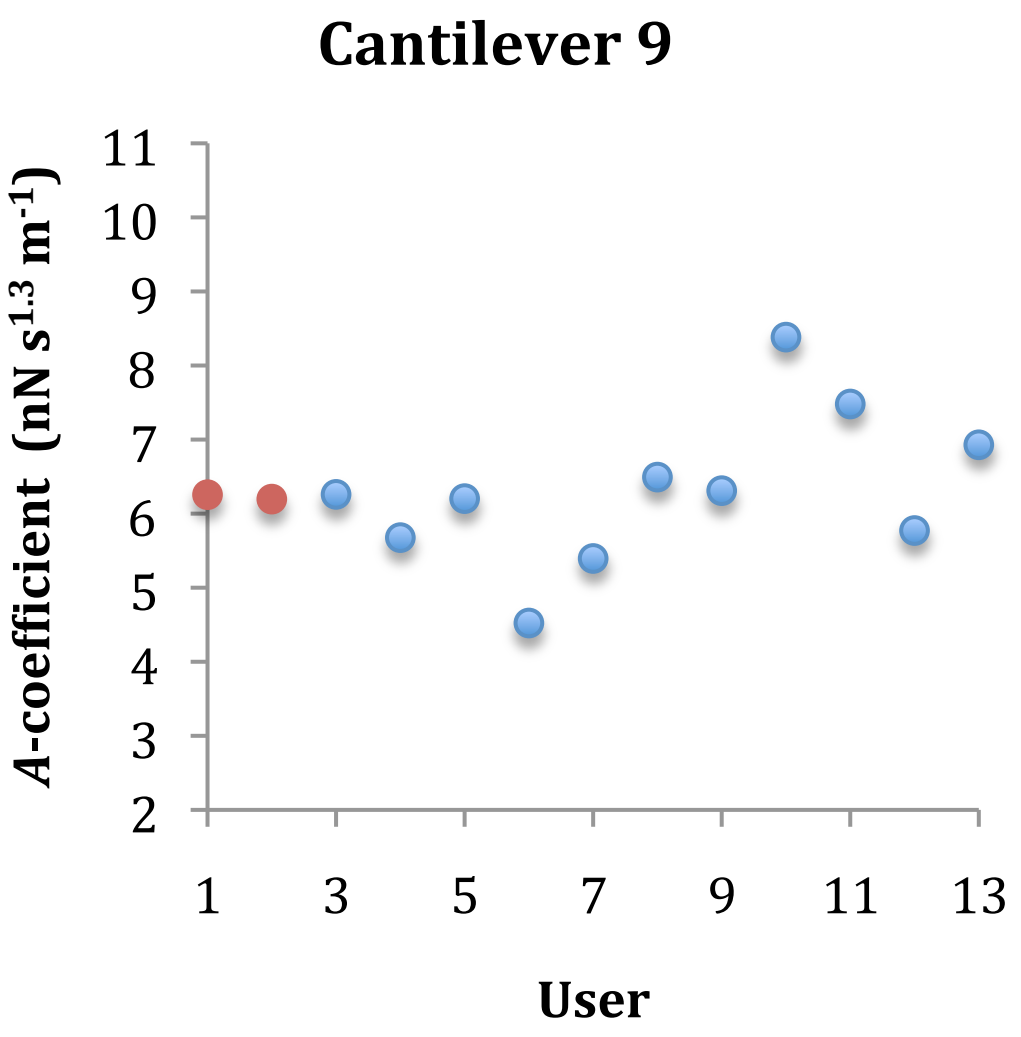

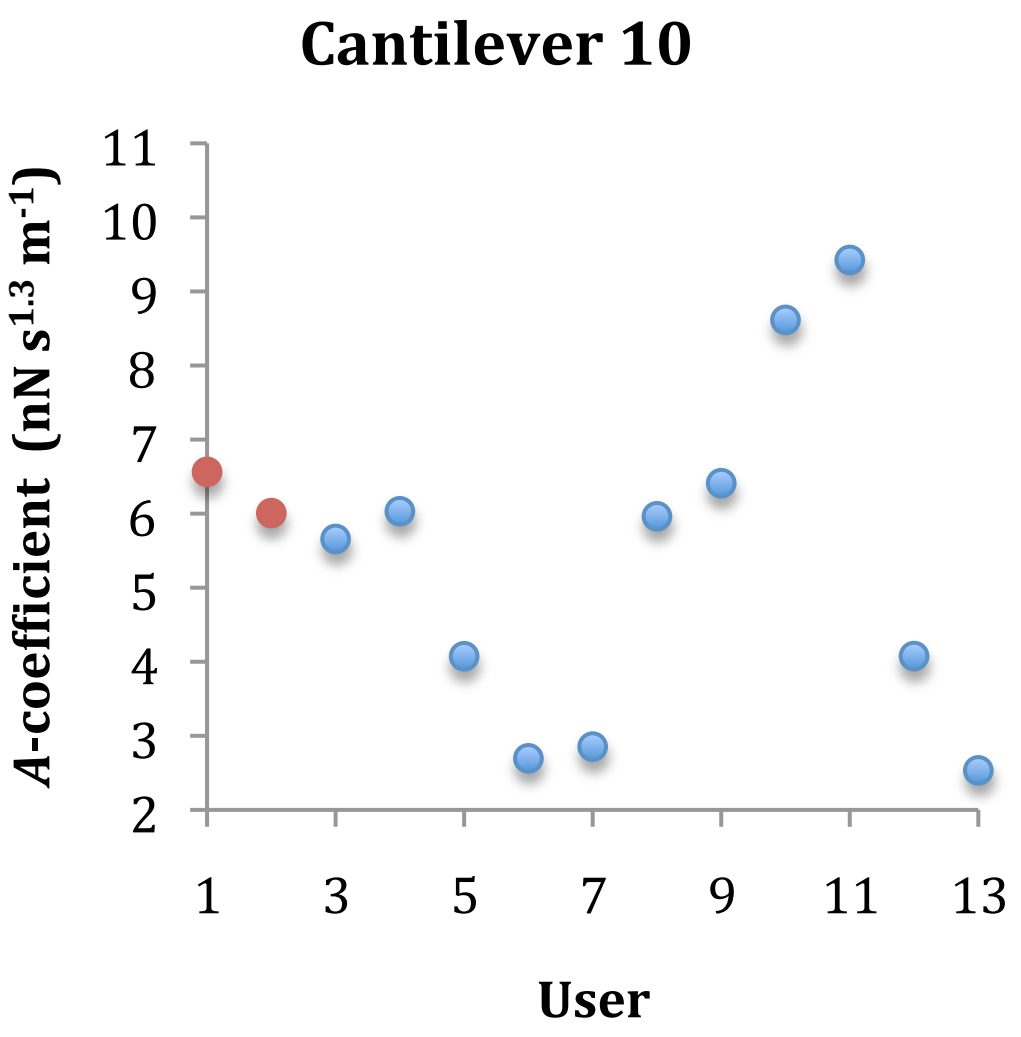

# Resonant frequency and quality factor dependence on tip-sample separation

Sample measurement of the quality factor and resonant frequency vs tip-sample separation, of an unknown AC240-R3 cantilever in air. Solid lines are measurement at a tip-sample separation of 3,000 μm, i.e., 100× the cantilever width (30 μm). The minimum separation measured is 0.75 μm.

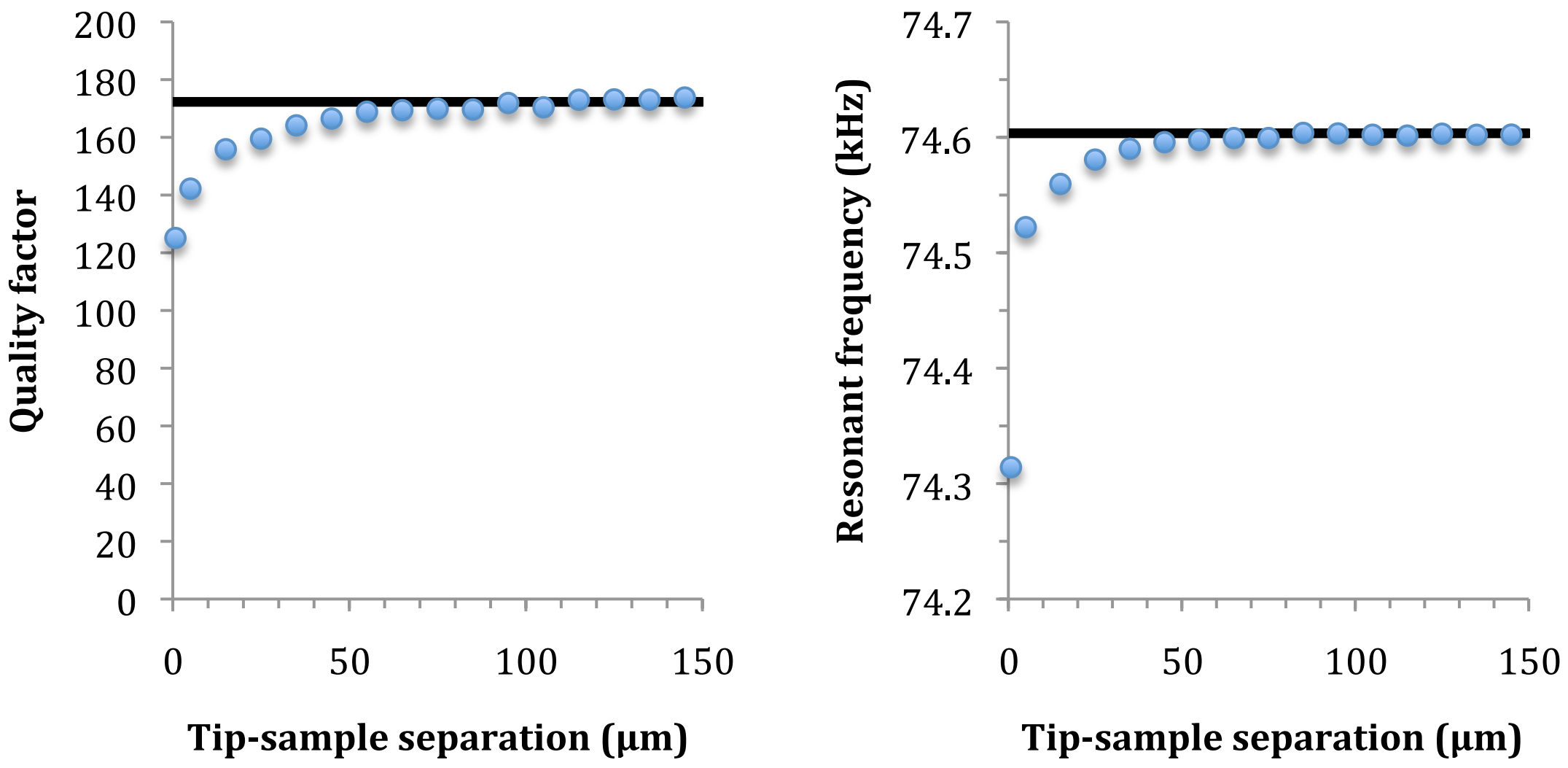

Quality factor is reduced significantly (~10-20%) for tip-sample separations less than 30 μm; this separation corresponds to the cantilever width. Operation at separations greater than two cantilever widths (60 μm) minimises this squeeze film damping effect.

While the resonant frequency also decreases with tip-sample separation, this reduction is two orders-of-magnitude smaller than observed for the quality factor.